# LIGHTCURVES OF 32 LARGE TRANSNEPTUNIAN OBJECTS

Susan D. Benecchi[1] & Scott S. Sheppard[1]

[1] Carnegie Institution of Washington, Department of Terrestrial Magnetism, 5241 Broad Branch Road, NW, Washington, DC 20015

**Proposed running head:** Lightcurves of Southern Hemisphere TNOs

**Corresponding author:** S. D. Benecchi, sbenecchi@dtm.ciw.edu; Carnegie Institution of Washington, Department of Terrestrial Magnetism, 5241 Broad Branch Road, NW, Washington, DC 20015 USA, Phone: 757-256-9751 or 202-478-8480, Fax: 202-478-8821

ABSTRACT

We present observations of 32 primarily bright, newly discovered Transneptunian objects (TNOs) observable from the Southern Hemisphere during 39 nights of observation with the Irénée du Pont 2.5-m telescope at Las Campañas Observatory. Our dataset includes objects in all dynamical classes, but is weighted towards Scattered objects. We find 15 objects for which we can fit periods and amplitudes to the data, and place lightcurve amplitude upper limits on the other 17 objects. Combining our sample with the larger lightcurve sample in the literature, we find a 3-sigma correlation between lightcurve amplitude and absolute magnitude with fainter objects having larger lightcurve amplitudes. We looked for correlations between lightcurve and individual orbital properties, but did not find any statistically significant results. However, if we consider lightcurve properties with respect to object dynamical classification, we find statistically different distributions between the Classical-Scattered and Classical-Resonant populations at the 95.60% and 94.64% level, respectively, with the Classical objects having larger amplitude lightcurves. The significance is 97.05% if the Scattered and Resonant populations are combined. The properties of binary lightcurves are largely consistent with the greater TNO population except in the case of tidally locked systems. All the Haumea family objects measured so far have measurable amplitudes and rotation periods ≤10 hr suggesting that they are not significantly different from the larger TNO population. We expect multiple factors are influencing object rotations: object size dominates lightcurve properties except in the case of tidal, or proportionally large collisional interactions with other TNOs, the influence of the latter being different for each TNO sub-population. We also present phase curves and colors for some of our objects.

*Subject headings*: **Planetary systems: Kuiper belt: general — Astronomical databases: Surveys**

***On-line material*: Table of photometry values; Magnitude vs. time figure for each object.**

## 1. INTRODUCTION

With discovery of the first Transneptunian object (TNO) in 1992 (Jewitt et al. 1993) a new area of study on small bodies in our solar system was opened. Most of these objects are located in the Kuiper Belt beyond the orbit of Neptune. The dynamically unstable Centaurs are found between the orbits of Jupiter and Neptune. More than 1600 of an expected $10^5$ objects (larger than ~100 km, Petit et al. 2011) have been observed and recorded in the Minor Planet Center (MPC) database. Of these, about half have orbits with small enough uncertainties to be observed from typical ground-based telescope facilities. Within the TNO population as a whole, objects are clustered in identifiable dynamical locations with respect to their interaction, or lack of interaction, with Neptune, or in areas near the plane of the Solar System (Elliot et al. 2005; Lykawka & Mukai 2007; Gladman et al. 2008). In this paper we use the classification system defined by the Deep Ecliptic Survey (DES; Elliot et al. 2005) and for statistical purposes combine objects in scattered orbits as described. In short, Cold Classical objects are TNOs with low eccentricity, circular orbits, inclinations less than 5-12° [Noll et al. 2008b; Elliot et al. 2005; Peixinho et al. 2008; for the statistical analyses in this paper we use 5.5°], and no previously traced interaction with Neptune. Resonance objects are in mean-motion resonances with Neptune. Scattered objects consist of Centaurs (objects with orbits inside of Neptune), Scattered disk objects (objects with large inclinations and eccentricities, with perihelia beyond 30AU that are not Resonant or Cold Classical objects), and Detached objects [objects with moderate to high eccentricities whose perihelia are sufficiently far (>40 AU) from Neptune so they are not influenced by Neptune]. Because of our broad grouping of objects for statistical analysis, the distinctions between the three classification systems is not significant.

The photometric variability of a TNO versus time, its lightcurve, is a powerful tool for learning about the shapes and surface features of these distant objects. Lightcurve amplitude, rotation period, color dependence, and shapes of lightcurves are all affected by the details the TNO's physical properties. The spin state of these bodies is important because it records the history of collisional and other evolutionary processes acting in the Kuiper Belt over time and at the extremes, can provide constraints on the material properties (and interior structure) of these objects. Disruption lifetimes for the largest objects (d≥400 km) are longer than the age of the Solar System, so we expect that the rotations of these objects are the results of impacts during the formation era of the Kuiper Belt (Lacerda 2005). Objects with d~200km have probably avoided catastrophic break-up, although their rotations could be modified by more recent collisions (Davis & Farinella 1997). The smallest objects (d≤100 km) are likely fragments resultant from multiple collisions over the age of the Solar System (Catullo et al. 1984; Lacerda 2005; see Campo & Benavidez 2012 for a review of the collisional environment spanning the range of TNO sizes). Likewise, TNO shape is also thought to be related to object size, with the largest objects being dominated by self-gravity and the smallest objects being collisional fragments without significant self-gravity influence.

In this paper we present lightcurve work carried out on 32 bright TNOs accessible from Las Campañas Observatory. Most of the objects in our target list were discovered as part of recent large area surveys conducted for bright outer solar system objects in the Southern Hemisphere (Sheppard et al. 2011; Rabinowitz et al. 2012). Other objects were selected from newly discovered TNOs in the MPC database with the requirement that they have absolute magnitudes, $H_V$, brighter than ~5.6 (corresponding to a size of d~330 km assuming an albedo of 0.1). We also observed binaries bright enough for our telescope-instrument combination ($m_R$<23.0). The survey includes five binaries and four Haumea family objects.

## 2. OBSERVATIONS

Observations were made using the Irénée du Pont 100" telescope at the Las Campañas Observatory during a series of eight 3-6 night runs (in 2007 and 2011-2012, Table 1). Images were collected using the direct CCD camera, a SITe2K chip with a pixel scale of 0.259 arcsec/pixel and a field of view of 8.85 arcmin square. The data were collected in pairs or triplets for 300 seconds in duration per exposure using the entire array and reading the chip in "fast" mode with a readout time of ~67 seconds. Observations for each TNO were collected on multiple nights within a single run with approximately 45-75 minute spacing between visits. The runs were scheduled in pairs 4-8 weeks apart so that lightcurve periods could be de-aliased from the 24 hour observing cycle within an observing run. Lightcurve data were collected using both the Sloan r' and Bessel R (Kron-Cousins equivalent) filters. Individual images have S/N of ≥30 (uncertainties ≤0.03 magnitudes), depending on the TNO magnitude and observing conditions. Photometric standard star data for overlapping stars from the Landolt (1992) and Sloan (Smith et al. 2002) surveys with TNO-like colors [104_428, 113_260, 113_339, RU_149B and RU_149F] were collected during photometric nights. We also collected colors for some of the TNOs during one photometric night of observation bracketing the Sloan g' and Sloan i' measurements with the r-band lightcurve filter (e.g., Sloan r' – Sloan g' – Sloan r' – Sloan i' – Sloan r') to ensure that variations did not unknowingly influence the colors. The conditions for each night are listed in Table 1.

INSERT TABLE 1 HERE

## 3. DATA ANALYSIS

For each run, a set of master biases, twilight sky flats and dome flats were generated to calibrate the data using standard routines from Dr. Marc Buie's IDL libraries (http://www.boulder.swri.edu/~buie/idl/). The data was flattened using sky flats for pointings with hour angles less than or equal to 1.0 hour East, and using dome flats for pointings with hour angles greater than 1 hour East; this cross-over point was empirically determined to provide the most consistent sky backgrounds across the chip. Each image was evaluated and the position of the TNO was identified by eye. Likewise, astrometry was carried out on each image, and on each star in the image with a peak pixel count <50,000 (the saturation level for the detector). For each object, three astrometric measurements (the first, last, and an observation in the middle of the sequence) were submitted to the MPC. Large (~4x the fwhm) and small (~fwhm) aperture photometry was carried out on all the stars identified on the image and saved to a file. Small aperture photometry was carried out on the TNO.

The star list was then culled in the following way to find the best comparison stars for each TNO: (1) The star was identified on each night of a single run, (2) the star was bright compared to the TNO (typically 3-4 magnitudes brighter than the TNO and with a peak pixel between 20,000-30,000 counts, well below the non-linear regime of the detector), (3) when plotted vs. time the star had no significant variations (the star was differenced by a selection of other stars of similar magnitude on the field to account for airmass and sky fluxuations, if the resulting difference varied by less then 3-sigma it was considered useable), and (4) the star was located relatively close to the TNO on the chip on at least one night of observation. After 10-15 comparison stars were identified each was evaluated by eye to ensure that it was not contaminated by a background source or itself a galaxy, typically a few of the stars were thrown out. Since TNOs move during the night relative to the background field stars these images were also examined vs. time to look for potential background contaminates. Images with background contamination due to faint galaxies, or nearby stars were eliminated. Most of the TNOs were in average density star fields, with the TNOs moving through clear areas, so few images were excluded (the fields were pre-selected as much as possible to avoid background object overlaps so this issue affects ≤5% of the data).

The selected field stars were photometrically calibrated using the standard star observations. If the sky was photometric for multiple nights in a given run, the calibrated star magnitudes were averaged across all the nights the field was observed. The scatter in the calibrated magnitudes was comparable to the photometric precision of the data and the calibrated magnitude uncertainties were small compared to the uncertainties on the TNO measurements themselves. To minimize uncertainties on the TNO photometry, the small aperture TNO magnitude for each frame was aperture corrected (Howell 2006) with the selected small and large aperture field star magnitudes. Finally the TNO magnitude was calibrated with a magnitude calibrated field star. For data collected in the Bessel R filter we apply a magnitude transformation of +0.202 magnitudes as determined using the *synphot* routine1 in IRAF (Laider et al. 2008) so that all values are based in the Sloan r' system.

In order to combine data across runs, the TNO magnitudes were geometrically corrected to unit distance from the Sun (*r*) and Earth (Δ) using the formalism $H_{r'}(\alpha) = R_{obs}(\alpha) - 5\log(r\Delta)$, where $\alpha$ is the phase angle of the observation, and $R_{obs}$ is the observed Sloan r' magnitude. In principle one would also want to correct the magnitudes to a phase angle of 0°, however, phase curves have not been measured for the majority of our objects and more than half of our objects span <0.3° in phase angle difference between measurements. However, for a few objects when we plotted all the data (magnitude vs. time) for that object, we noticed some large-ish (0.2-0.3 magnitude) offsets from earlier runs which observed the same objects (primarily data from the May 2012 dataset which was collected under constantly varying conditions); some of these offsets we thought could be explained by geometry, others could not. Therefore, we period-fit our data considering three different magnitude calculations: (1) geometric correction excluding the phase coefficient as expressed earlier in this paragraph, (2) applying the Bowell et al. (1989) formalism whose equations are given in the table note of Table 2, and (3) normalizing the values on each night by the mean magnitude of the object.

## 4. LIGHTCURVE RESULTS

### *4.1. Analysis*

Table 2 provides detailed observational and geometric circumstances for each TNO on each night of observation; the full table is provided electronically. It records the mid-time (in both calendar format and JD) of the observations, the magnitude range of the observations, the

1 In *synphot* we use the Kurucz solar model as our reference standard and a reddening function (ebmv) of 0.1 which corresponds to a color term of ~0.55 magnitudes in V-R (Benecchi et al. 2011).

number of observations, the duration of time (hours) covered by the observations, heliocentric distance (AU), geocentric distance (AU), phase angle (°), light time (s) and two magnitudes [$H_{R,Bowell}$ which uses the Bowell et al. 1989 formalism and $H_R(\alpha)$, the reduced magnitude uncorrected for phase angle] for each object, each night. Table 3 provides basic characteristic information about each object and a summary of our lightcurve conclusions from all of our observations combined. We ran the dataset for each object through a set of periodogram analyses (described in the following paragraph), for periods ranging from 3 to 20 hours based on our observing windows and aliasing considerations. Some of the resulting phasing or suggested periods were not statistically significant, and instead of making a "best guess" which might be carried forward in the literature incorrectly, we choose instead to place an upper limit on the amplitude of a possible lightcurve for the duration of a single night of observation (≤8 hours) by averaging the range on all nights of observation. Individual uncertainties on the measurements are ~0.03 magnitudes or better. Individual measurements and plots of the data vs. time on each night of observation can be found in online supplemental materials. We include in Table 4 and Figure 1 samples of the online data table and figures which include values for all our objects, regardless if they resulted in conclusive period fits, or flat lightcurves.

INSERT TABLE 2, 3, 4 AND FIGURE 1

ONLINE TABLES AND FIGURES

For objects with data to which we can fit periods with statistically significant results, we present in Figure 2 the Lomb-Scargle periodogram (top; Scargle et al. 1982) and the data phased to the most significant period assuming single- (middle) and double- (bottom) peaked lightcurve interpretations. For each dataset we calculate the 99.9% confidence level using the methods described by Horne & Baliunas (1986) and plot the result as a dashed line on each of the periodogram plots. Very similar results are found when using the phase-dispersed minimization (PDM, Stellingwerf 1978) and Harris-Foster (Harris et al. 1989; Foster 1995) fitting techniques. For objects with multiple peaks above the confidence interval we phased the data to each possible period for evaluation, but in all cases the resulting period was the one with the highest Lomb-normalized spectral power value. We list in the notes column of Table 3 possible additional periods that could fit the data (that we were unable to completely rule out), although we run all further analyses with our preferred periods, listed in the specified period column. We fit all three sets of magnitude calculations for each object. We draw our conclusions based on the night normalized data which minimizes any offsets between runs, although it could introduce a bias against objects observed for short periods of time with long period >8 hour rotation curves. Fits on the non-phase-corrected and Bowell-corrected values were also carried out and, where offsets were not an issue, the same periods were identified. Offsets were on order 0.1-0.2 magnitudes, which are within the range of what one would expect for phase corrections. However, we did not sample a complete range of phases for all our objects, only for some objects (see section 6). For consistency we analyze all the lightcurves using the night normalized data.

INSERT FIGURE 2 HERE

### *4.2. This Sample*

Our dataset includes objects in all dynamical classes2 (Table 1 provides a numerical summary). We find fifteen objects for which we can fit periods to the data (~46% of our sample); eight of these are from the Scattered population, one is a Haumea family object, one is Classical, and five are Resonant objects.

Our sample includes eight objects with amplitudes ≥0.2 magnitudes. Two of the objects with the largest variations are unresolved Cold Classical binaries and the smallest/faintest objects in our sample [2005 $EF_{298}$ and (303712) 2005 $PR_{21}$]. One of these, 2005 $EF_{298}$ is best fit with a 4.82 hour single-peaked or 9.65 hour double-peaked period although a period of 6.09/12.18 hours also gives a decent result. We do not have enough data to constrain the period for (303712) 2005 $PR_{21}$. 2010 $VK_{201}$, a Scattered object, is fit with a period of 3.79 or 7.59 hours and 2007 $JF_{43}$, a 3:2 Resonant object, is fit with a period of 4.76 or 9.52 hours. 2009 $YE_{7}$, a Haumea family object (Trujillo et al. 2011), is variable at ~0.2 magnitudes, but we were not able to fit a unique period to our dataset. 2010 $FX_{86}$ is clearly variable (~0.2 magnitudes or greater) within individual nights of observation, though, a number of periods are significant above the confidence level including the one we choose at 7.90 hours (single-peaked) or 15.8 hours (double-peaked) for our analyses. (307251) 2002 $KW_{14}$ was fit with a preferred period of 6.63 hours and an amplitude of 0.25±0.03 magnitudes. The 4.29/8.57 hour and 5.25/10.5 hour period peaks for (307251) 2002 $KW_{14}$ from Thirouin et al. (2012) are still possible, however only the later period fits above our confidence level; our lightcurve amplitudes of 0.25±0.03 and (0.21 or 0.26)±0.03 magnitudes, respectively, are in agreement. 2010 $PU_{75}$ is a Scattered object which we fit with a rotation period of 6.19 or 12.39 hours and an amplitude of 0.27±0.03 magnitudes.

We are able to fit periods for another nine of our objects, all with variations <0.18 magnitudes. The best-fit periods range from ~3.2 hours to 19.5 hours, considering both single- and double-peaked lightcurve interpretations. Three of these objects, 2010 $EL_{139}$, 2010 $EK_{139}$, and 2010 $ET_{65}$ have single-peaked rotation periods <4 hours. 2010 $EK_{138}$ is the largest object of this set assuming comparable albedos. The double-peaked lightcurves at 6.32 hours, 7.07 hours and 7.88 hours, respectively, also give reasonable results. These are good candidates for Jacobi ellipsoids, elongated by their fast spins (Jewitt & Sheppard 2002); they all have lightcurve amplitudes ~0.13 magnitudes. 2007 $JJ_{43}$ and (312645) 2010 $EP_{65}$ were both fit with preferred single-peaked periods of 6.04 and 7.48 hours, respectively. Both objects are equally well interpreted as double-peaked with periods of 12.08 and 14.97 hours and could also be fit with periods about an hour shorter or longer than the chosen periods. (303775) 2005 $QU_{182}$ and 2010 $HE_{79}$ are both fit with single-peaked periods just shy of 10 hours (9.61 and 9.75 hours). Both are likely large enough objects to have their shapes dominated by gravity.

Our preferred period, 6.95/13.89 hours (single/double peaked), for (145452) 2005 $RN_{43}$ is different than the preferred period of Thirouin et al. 2010. Both of their single-peaked periods: 5.62 and 7.32 hours are nominally consistent with our dataset, but neither period fits our dataset above the confidence level. Our amplitudes are in agreement, 0.04±0.01 and 0.06±0.01 magnitudes, respectively. We fit a period of 4.85 or 9.71 hours to (120178) 2003 $OP_{32}$, with an amplitude of 0.18±0.01 magnitudes, slightly longer then the interpretation of Thirouin et al.

2 Using the DES classification system. This can be found for each object at the following address where "objectname" is the MPC designation (number or preliminary designation dropping the first two numbers of the year, eg. 05EF298): http://www.boulder.swri.edu/~buie/kbo/astrom/05EF298.html

2010; their preferred period of 4.05 hours is not consistent with our dataset although their lightcurve amplitude of 0.13±0.01 magnitudes is comparable.

One object in our sample is consistent with a long period interpretation, (79360) Sila/Nunam. This object is both a Cold Classical TNO and a binary currently undergoing mutual occultations and eclipses as reported in Grundy et al. (2012). Our 2011 dataset was included in Grundy et al. (2012). We include in this data sample our 2012 observations. We don't observe enough of the period to fit the data to the mutual orbital period of the components, but our data are consistent with such an interpretation. Our amplitude variation of <0.17 magnitudes is also consistent with the 0.14±0.07 magnitude variation reported by Grundy et al. (2012).

### *4.3. Axis Ratio*

In addition to rotation period, we can use the amplitude of variation to estimate the sphericity of our objects. Our object sample ranges from $H_V$=2-6.1 (or d~1600-250 km using the formalism of Bowell et al. 1989 and assuming an albedo of 0.1), and includes objects from both the spherical and elongated groups. We expect the smaller objects to be elongated, having double-peaked lightcurves where nominally we can sample the long and short axes of the object twice over a full rotation. If we assume such an object to be triaxial with semi-major axes a≥b≥c in rotation about the c-axis, the minimum and maximum flux of the rotation curve measured in magnitudes, Δm, can be used to determine the projection of the body shape (i.e. how spherical the object is) into the plane of the sky:

$$\Delta m = 2.5\log\frac{a}{b} - 1.25\log\left(\frac{a^2\cos^2\theta + c^2\sin^2\theta}{b^2\cos^2\theta + c^2\sin^2\theta}\right) \quad (1)$$

where θ is the angle at which the rotation axis is inclined to the line of sight (an object with θ=90° is being viewed equatorially; Binzel et al. 1989). If we assume that we are in fact viewing the object equatorially, then this equation can be rearranged to give the axis ratio, $a/b = 10^{0.4\Delta m}$. Our sample ranges from 1.03-1.33 with an average of 1.16.

### *4.4. Previous Results*

The existing database for the rotation properties of TNOs (including Centaurs) is on the order of 100 objects. About two dozen objects have been studied in extensive detail while many have only been observed in one or two observing campaigns. Observations generally require substantial amounts of telescope time on 2-m class or larger telescopes for the brightest objects ($m_R$≤23.0), and time on 4-m class or larger telescopes for the fainter population of objects. A variety of studies have been done to estimate typical lightcurve amplitudes and periods, each study having its own brightness and size limitations. Numerous biases also exist among the datasets: (1) objects with longer periods are harder to observe due to a combination of the fact that objects are up no more than ~8.5 hours above 30° (an airmass of 2) at most observing locations if only one telescope facility is used, (2) faint and/or small objects are not possible to observe with 2-m class telescopes, the ones that are easier to get long consecutive stretches of telescope time on, (3) most of the brightest objects are in the Plutino (3:2 resonance) and Scattered disk population, so studies of objects by dynamical classification requires access to a combination of telescope sizes and facilities, and (4) observations over a few nights on a single observing run might not result in a unique object period.

The recent summary paper of Duffard et al. (2009) compiles 91 lightcurves (74 TNOs and 17 Centaurs) from the literature and their own work (inclusive of Thirouin et al. 2010) to conclude that the mean rotation period for all TNOs is 7.35 hours, or 7.71 hours for the TNOs alone, excluding Centaurs. This sample includes objects from all dynamical regions of the Kuiper Belt, as well as a handful of unresolved binary objects. Except for the Centaur population, small sample sizes have limited our ability to investigate correlations between lightcurves and dynamical properties; however, the number of objects with measurements is now becoming large enough to consider such correlations.

One difficulty of summary analyses for lightcurves is if an object has been observed, but no unique rotation period has been identified. In some cases a number of periods are equally plausible in the absence of more data, and all are recorded in the literature. For summary analyses Monte Carlo models can be used to include these results; however, uniquely determined periods are preferable. Another complication is how the lightcurve is interpreted, as a single- or double-peaked curve. The lightcurve of an elongated TNO will be due to changes in the projected cross-section of the object as measured at the telescope. The lightcurve of a spherical TNO, presumed not to have an atmosphere, is most likely to be caused by surface variations in either albedo or topography across the surface of the object. In this case, lightcurve amplitudes are typically small and, based on asteroid studies, are empirically found to be less than 10-20% (Magnusson, 1991). For TNOs it has been suggested that an amplitude of ~0.15 magnitudes is a reasonable break point for interpreting lightcurve amplitudes as being due primarily to surface albedo variations (≤0.15 magnitudes) or due to object elongation (>0.15 magnitudes; Sheppard et al. 2008; Thirouin et al. 2010]). As the number of good lightcurves grows and as other methods of shape determination are employed (e.g. occultations, Person et al. 2006), we may be able to better refine this distinction for TNOs as a whole, or for different subsets of the population.

## 5. DISCUSSION

In the following discussion we include results from this work and that collected from the literature out of the references listed in the caption of Figure 3. We use absolute magnitudes $H_V$ from the MPC so as to interpret all objects in a consistent manner relative to intrinsic brightness which can be used as a proxy for size. One method of estimating an effective diameter for these objects is to follow the formalism of Bowell et al. (1989), where $d = 10^{((6.259-0.4*H_V-\log\rho)/2)}$ km and $\rho$ is albedo. It is known that TNO albedos range from ~0.04 to ~0.8 (Stansberry et al. 2008). Because the objects in our sample are relatively new discoveries, they do not yet have measured albedos. For our calculations we work in $H_V$ space, but to give a general reference point on size for Figures 3 and 5 we assume $\rho$=0.1. Figure 3 shows a histogram of absolute magnitude for the entire sample of published TNO lightcurves sub-divided by dynamical classification, with the sample for this work extracted in a separate plot. The majority of objects sampled thus far are from the dynamically Scattered population since these objects tend to be intrinsically bright and more easily observed from smaller telescopes.

In Figure 4 we plot the cumulative number of objects with respect to lightcurve amplitude normalized by the total number of objects in the sample for the entire measured population and also for the sample in this work alone using our fitted amplitudes or upper limits. The solid points/black line shows the results for the full sample, the triangles/red line for the Classical objects, the squares/gray line for the Resonant objects and the diamonds/blue line for the Scattered objects. In both samples the Scattered objects typically have smaller amplitude lightcurves then the larger population. However, Scattered objects at all sizes are measured, so size is not the sole explanation for this effect.

In Figure 5 we plot rotation period and amplitude vs. absolute magnitude, and find that objects fainter than about $H_V$~5 appear to have larger amplitude lightcurves. Using the entire sample, a Spearman rank correlation test (Table 5) between absolute magnitude and lightcurve amplitude indicates this to be a 3-sigma result. We also see an indication of correlation between absolute magnitude and single period rotation curves, although this is less than a 3-sigma result and there is also some ambiguity between single and double peaked lightcurve interpretations. We consider correlations with orbital properties, but do not find any statistically significant results. If we run the same analysis for only the binaries or Haumea family objects (discussed in sections 4.4 and 4.5), we also do not find significant correlations. It appears that size has a greater influence over the rotational properties of an object than one particular orbital characteristic, unless there has been obvious interaction with another TNOs as in the case of tidally locked binaries.

Figure 5 also demonstrates that, with the exception of tidally locked objects, the rotation rate for the majority of measured objects is less than 13/26 hours (single-peak/double-peak interpretations), with the mean rotation periods being 6.73 hours and 11.30 hours, respectively (excluding Pluto/Charon and Sila/Nunam). The scatter is relatively small. Modeling by Lacerda (2005) suggests that such a spin distribution is indicative of some level of anisotropic accretion in the early Kuiper Belt.

INSERT FIGURE 3, FIGURE 4, AND FIGURE 5 HERE

INSERT TABLE 5 HERE

However, if we compare the rotational properties of objects to each other with respect to dynamical class, binary, or family status (Table 6) we find a 95.60% probability that the Classical and Scattered object amplitudes come from different distributions and a 94.64% probability that the Classical and Resonant object amplitudes come from different distributions. If we combine the Scattered and Resonant objects the significance of the difference increases to 97.05%. We also investigated if our result was significantly affected by the absolute magnitude range of the sample. We could not test for samples brighter than an absolute magnitude of 6.5 because the sample size for the Classical objects is too small, however, we plot in Figure 6 the % results vs. absolute magnitude for samples from 6.5 to 12 (the faintest object) in steps of 0.5 magnitudes. The amplitude distinction we find is strongest for all three samples (Classical/Resonant; Classical/Scattered and Classical/Scattered+Resonant) with an absolute magnitude limit brighter than 7.0, however, it is strong in all magnitude bins and we believe that the effect is real, not size-dependent. So while one particular orbital element does not dominate, it does appear that location within the belt plays a significant role in rotation properties. Perhaps objects in more stirred-up regions have experienced greater interaction with other bodies, but the amplitudes of larger objects are disproportionately affected by smaller collisions. There are not enough objects to compare within the binary or family populations themselves. We present all the correlations we considered in Table 6 for completeness.

INSERT TABLE 6 HERE

INSERT FIGURE 6

### *5.1. Binaries*

Binaries are found throughout the Transneptunian belt, though in significantly higher fractions among the Cold Classical population [ $29^{+7}_{-6}\%$ vs. $5.5^{+4}_{-2}\%$ for all other classes combined (Noll et al. 2008a,b)]. A number of formation mechanisms for these systems have been proposed including: (1) physical collisions (Weidenschilling 2002; Canup 2005 & 2011), (2) gravitational interactions (Goldreich et al. 2002; Astakhov et al. 2005; Lee at al 2005; Funato et al. 2003), and (3) gravitational collapse (Nesvorny et al. 2010). Each of these mechanisms has the potential to influence the rotational properties of these objects, in addition to the tidal and orbital interactions between the binary objects themselves. Tidal interactions can have the effect of synchronizing the rotation period of an object with the mutual binary orbit, as is well known in the case of the (134340) Pluto/Charon system (Tholen & Tedesco 1994). Grundy et al (2012) and this work also support the interpretation that (79360) Sila/Nunam is a tidally locked system. There are currently 24 binaries with lightcurve measurements and/or estimates.

Our sample includes five binary objects and, as mentioned in section 4.1, two of these objects have the largest variations and are the smallest objects in our sample. The two binaries that do not show significant variation are the brightest of our binary sample, consistent with the idea that smaller objects have larger amplitude lightcurves. In Figure 7 we plot the lightcurve characteristics of all the binaries in the literature (references can be found in the figure caption). In a statistical sense, the sample is still small, however we note that the same low amplitude characteristic of the Scattered objects is seen. The Cold Classical objects have the largest amplitudes with the exception of the resonant object 2001 $QG_{298}$, whose large amplitude lightcurve is consistent with a contact binary interpretation (Sheppard & Jewitt, 2004; Takahaski & Ip, 2004). We find hints that the binary amplitudes as a whole may be slightly larger then the non-binary population, but overall the distributions are similar and we are hesitant to over-interpret the statistics.

INSERT FIGURE 7 HERE

### *5.2. Haumea Family Objects*

In the asteroid belt, much has been learned through study of asteroid families with respect to both dynamical and photometric properties. Modeling the dynamics of collision family members in many cases can trace back a timeframe for when family creation occurred (Nesvorny et al. 2006). In the Koronis family, spin studies of large numbers of objects of various sizes (Slivan 2002; Slivan et al. 2008, 2009) have demonstrated markedly nonrandom alignments of spin obliquities and correlations with spin rates which in the asteroid belt are interpreted to be thermally driven (Vokrouhlický et al. 2003). The modification of these spin rates is the result of the Yarkovsky (YORP) effect which disproportionately heats non-spherical objects and has the effect of increasing rotation rates of objects; this effect is stronger the smaller the object. YORP is not effective in the Kuiper Belt; since objects are too distant from the Sun.

In the Kuiper Belt one dynamical family has been identified through spectroscopic studies (Barkume et al. 2006), and confirmed with dynamical integrations (Ragozzine & Brown 2007). Rotational studies of the largest body, (136108) Haumea, found it to have a rapid rotation, 3.9154±0.0002 hours (double-peaked) with an amplitude of 0.28±0.04 magnitudes, which can be explained as the result of a physical collision (Rabinowitz et al. 2006, Schlichtling et al. 2009, Leinhardt et al. 2010; Lykawka et al. 2012), although Ortiz et al. (2012) argue that such a system could also be created through rotational fission due to collisional spin-up. Studies of the rotation properties of the Haumea family may provide insight for spin properties resultant from a formation mechanism independent from modification by thermal factors. It is possible that small collisions play a roll in rotational modification, but to date this has not been demonstrated. It is believed that large collisions influence the spin properties of the target and the material ejected during such an event (Paolicchi et al. 2008). If the Haumea system is the result of a collision, one might expect the spin properties of resultant family members to be different as a group from the background TNO lightcurve distribution. Perhaps these objects are more elongated as a result of the collision, or spinning more rapidly (small objects) or more slowly (large objects), depending on the energy of the initial collision.

The numbers are still too small for statistics and many of the smaller objects still need to be studied; however, we present in Figure 8 the current lightcurve characteristics of 9 objects from this work and the literature. All have rotation periods (where measured) ≤10 hr and distinguishable light curve amplitudes. The mean single-peaked period is 5.60 hours and the mean double-peaked period is 7.39 hours inclusive of Haumea itself. The mean amplitude is 0.14 magnitudes. These values are in comparison with the greater TNO population (Centaurs included, Pluto/Charon and Sila/Nunam excluded) with has a mean single-peaked period of 6.73 hours, a mean double-peaked period of 11.30 hours and a mean amplitude of 0.20 magnitudes. At this point, only the brightest Haumea family objects have been observed. Since all the amplitudes are small, it is likely that they can be interpreted as spherical objects with surface variations due to albedo features. However, both Haumea and (55636) 2002 $TX_{300}$ are known to have high albedos compared to other TNOs (Lellouch et al. 2010; Elliot et al. 2010; Mommert et al. 2012). If all Haumea family members have high albedos then the objects would be smaller and perhaps these amplitudes are due to object elongation.

INSERT FIGURE 8 HERE

## 6. PHASE CURVE AND COLOR RESULTS

A phase curve describes the brightness of a TNO as a function of its phase angle, the angle made between the Sun, the TNO, and the observer (Earth); for TNOs, the maximum angle is ~2°. It is linear outside of a few tenths of degrees, but studies of asteroids and moons of the giant planets find non-linear brightening as the phase gets close to zero (Verbiscer et al. 2005). The surge can be explained by two physical mechanisms, shadow hiding and coherent backscattering; both are related to what is happening at the surface of the TNO. Shadow hiding is the result of hills, boulders or a mix of light and dark ices on the surface of the object. At low phase angles, no shadows occur and the object appears brighter than at larger phase angles where shadows contribute to the disk-integrated photometry. Coherent backscattering occurs when multiply scattered rays bounce off the surface of the object and follow the same path back to the observer; the light rays add together and a brightening occurs. Near zero-phase angle light paths interfere more constructively as seen by the observer than at larger phase angles (Schaefer el al. 2009).

Slightly less than one third of our objects span a large enough region of phase angle space (≥0.3°) for us to estimate the linear phase coefficient, which can be expressed in flux as $\phi(\alpha) = 10^{-0.4\beta\alpha}$, where β is the phase coefficient in magnitudes per degree at phase angles, $\alpha<2°$. Figure 9 plots our measurements and Table 7 gives the results of our fit for each object. We find an average phase coefficient of $\beta_R$= 0.23 mag/° for the objects we can measure (with $\alpha>0.2°$), higher than that found by Belskaya *et al. (*2003), although consistent with some of the individual values listed in Rabinowitz et al. (2007) & Schaefer et al. (2009). (278361) 2007 $JJ_{43}$ which has the steepest slope, and relatively small uncertainty, is measured near a phase angle of 0.2°, close to the region where the opposition surge can have an effect. We don't have any measurements between the two extremes so we suggest these values be used with caution. We do not have albedo and color-phase measurements for these objects, but based on the criteria established in Schaefer et al. (2009) for the phase curve slope, we infer that if we are seeing a surge effect it is most likely due to the coherent backscattering mechanism.

INSERT FIGURE 9 HERE

INSERT TABLE 7 HERE

Additionally, we collected color information on one night for each object. The individual measurements can be found in Table 8 and the cumulative results are plotted in Figure 10. Our objects span the range of TNO colors and are redder than the Sun with the exception of 2009 $YE_7$, which is known to be a Haumea family member (Trujillo et al. 2011). None of the other objects for which we measured colors are extreme compared to other objects in the Kuiper Belt (Sheppard 2010; Benecchi et al. 2011; Fraser et al. 2012).

INSERT FIGURE 10 HERE

INSERT TABLE 8 HERE

## 7. CONCLUSIONS

We have presented lightcurve results for 32 large Southern Hemisphere TNOs. We can fit rotation periods with statistically significant results for 15 of these and place amplitude limits on the lightcurves of 17 other objects. All of the objects in our sample have amplitudes ≤0.3 magnitudes with periods ≤10 hours (assuming a single-peaked interpretation) with the exception of (79360) Sila/Nunam for which our data is consistent with the mutual orbit binary period. We find an average axis ratio of 1.16 suggestive of albedo variations on these objects. Combining lightcurve results from the literature with our measurements we find a correlation between lightcurve amplitude and absolute magnitude at the 3-sigma level, with small objects having larger amplitudes. However, the correlation is not as statistically significant if one considers the binary or Haumea family populations on their own. We suggest that size has a greater influence over the rotational properties of an object than one particular orbital component. We also find that comparison of lightcurve amplitude with respect to dynamical population results in statistically different distributions between the Classical/Scattered and Classical/Resonant populations at the 95.60 and 94.64% levels, respectively, with the Classical objects having higher amplitude lightcurves. This statistic is 97.05% if the Scattered and Resonant parameters are combined. Perhaps multiple factors are at work, objects in more stirred-up regions (characteristics of binaries demonstrate that the cold classical region has been significantly less perturbed than the Scattered and Resonant populations in particular with respect to the number of binary systems and their orbital properties; Grundy et al. 2011; Parker et al. 2011) have experienced greater interaction with other bodies and have likewise had their shapes more greatly altered. Large bodies are less susceptible to large scale changes unless collisions are of comparable sized objects. The properties of binary lightcurves are largely consistent with the greater TNO population except in the case of tidally locked systems. The nine Haumea family objects with measured light curves have rotation periods ≤10 hr; none have completely flat light curves. It is likely that all of these objects are large enough to be dominated by surface albedo variations; alternatively, if all Haumea family objects have high albedos, then the objects are actually smaller and the lightcurve properties would be dominated by shape.

ACKNOWLEDGMENTS

This paper includes data gathered with the 100" Irénée du Pont telescope located at Las Campañas Observatory, Chile operated by the Carnegie Institution of Washington. We wish to thank telescope operators Herman Olivares, Sergio Castellón and Nidia Morrell and telescope support staff Patricio Pinto and Oscar Duhalde. The work for this paper was supported through a Carnegie Fellowship at the Department of Terrestrial Magnetism. We also thank an anonymous reviewer for helpful comments that improved the manuscript.

## TABLES

TABLE 1. DU PONT RUN DETAILS

| Run Dates (UT) | Observing Conditions | Filter | Objects Observed |
|---|---|---|---|
| 2007 July 15 | photometric | Bessel R | 119951, 120178, 120347, 145452, |
| 2007 July 16 | photometric | | 145453, 174567, 307251, 2007 $JH_{43}$ |
| 2007 July 17 | photometric | | |
| 2007 July 18 | photometric | | |
| 2007 July 19 | photometric | | |
| 2007 July 20 | photometric | | |
| 2007 August 15 | photometric | Bessel R | 120178, 307251, 119951, 145453 |
| 2007 August 16 | photometric | | |
| 2007 August 17 | photometric | | |
| 2011 March 09 | photometric | Sloan r' | 79360, 278361, 312645, 2010 $ER_{65}$, |
| 2011 March 10 | photometric | | 2010 $ET_{65}$, 2010 $FX_{86}$, 2010 $EK_{139}$, 2010 |
| 2011 March 11 | photometric | | $EL_{139}$ |
| 2011 March 12 | photometric | | |
| 2011 March 13 | photometric | | |
| 2011 March 31 | photometric | Sloan r' | 278361, 312645, 2007 $JF_{43}$, 2010 $ER_{65}$, |
| 2011 April 2 | photometric | | 2010 $FX_{86}$, 2010 $HE_{79}$, 2006 $HJ_{123}$, |
| 2011 April 3 | photometric | | 2010 $EK_{139}$, 2010 $EL_{139}$ |
| 2011 April 4 | photometric | | |
| 2011 April 5 | photometric | | |
| 2011 September 28 | photometric | Sloan r' | 225088, 2009 $YE_{7}$, 2008 QY40, |
| 2011 September 29 | photometric | | 2010 $RF_{43}$, 2005$QU_{182}$, 2010 $VK_{201}$ |
| 2011 September 30 | photometric | | |
| 2011 October 1 | photometric | | |
| 2011 October 2 | photometric | | |

| Run Dates (UT) | Observing Conditions | Filter | Objects Observed |
|---|---|---|---|
| 2011 October 20 | photometric | Sloan r' | 225088, 303712, 2008 $QY_{40}$, 2010 $RF_{43}$, |
| 2011 October 21 | photometric | | 2010 $RO_{64}$, 2010 $TY_{53}$, 2010 $VZ_{98}$, |
| 2011 October 22 | photometric | | 2005 $QU_{182}$, 2010 $VK_{201}$ |
| 2011 October 23 | clouds, high wind | | |
| 2011 October 24 | clouds, high wind | | |
| 2012 March 18 | photometric | Bessel R | 79360, 278361, 312645, 2007 $JH_{43}$, |
| 2012 March 19 | photometric | | 2010 $ET_{65}$, 2010 $FX_{86}$, 2010 $HE_{79}$, 2010 |
| 2012 March 20 | photometric | | $KZ_{39}$, 2005 $EF_{298}$ |
| 2012 March 21 | photometric | | |
| 2012 March 22 | photometric | | |
| 2012 May 13 | variable thick clouds | Sloan r' | 2007 $JF_{43}$, 2007 $JH_{43}$, 278361, 2010 $ER_{65}$, |
| 2012 May 14 | high winds | | 2010 $ET_{65}$, 2010 $FX_{86}$, 2010 $PU_{75}$ |
| 2012 May 15 | high winds | | |
| 2012 May 16 | photometric, high winds | | |
| 2012 May 18 | clouds by night end | | |

**Object Classification Summary**

| Classification | Number of Objects in Our Sample | Notes |
|---|---|---|
| Cold Classical | 4 | All but one is binary[a] |
| Resonant | 7 | |
| Scattered | 17 | Inclusive of Centaurs; 4 Haumea Family[b] |

[a] Noll et al. 2006; Stephens & Noll 2006; Noll et al. 2008c; Noll et al. 2011.

[b] Ragozzine & Brown 2007; Trujillo et al. 2011.

TABLE 2. DU PONT OBSERVATION DETAILS

| Object | Calendar Date[a] | JD (mid-time)[a] | $\Delta m^b$ | $N_{obs}$ | $\Delta t^c$ (hr) | r (AU) | Δ (AU) | α (°) | L-Time (s) | $H_{r',Bowell}{}^d$ | $H_{r'}(\alpha)^e$ |
|---|---|---|---|---|---|---|---|---|---|---|---|
| 79360 | 2011 March 09 02.09829 | 2455629.58743 | 0.12 | 8 | 2.57 | 43.501 | 42.704 | 0.786 | 355.152 | 5.10 | 5.23 |
| 79360 | 2011 March 10 01.77201 | 2455630.57383 | 0.11 | 8 | 2.47 | 43.501 | 42.714 | 0.804 | 355.237 | 5.18 | 5.31 |
| 79360 | 2011 March 11 01.70889 | 2455631.57120 | 0.18 | 8 | 2.59 | 43.501 | 42.724 | 0.823 | 355.324 | 5.15 | 5.28 |
| 79360 | 2011 March 12 01.21948 | 2455632.55081 | 0.17 | 6 | 1.44 | 43.501 | 42.735 | 0.841 | 355.413 | 5.11 | 5.25 |
| 79360 | 2011 March 13 01.62795 | 2455633.56783 | 0.17 | 8 | 2.43 | 43.501 | 42.746 | 0.859 | 355.505 | 5.00 | 5.14 |
| 79360 | 2012 March 18 01.64508 | 2456004.56855 | 0.28 | 6 | 2.35 | 43.496 | 42.793 | 0.935 | 355.893 | 4.94 | 5.09 |
| 79360 | 2012 March 19 02.03244 | 2456005.58469 | 0.05 | 6 | 3.62 | 43.496 | 42.805 | 0.951 | 355.993 | 5.10 | 5.25 |
| 79360 | 2012 March20 01.14035 | 2456006.54751 | 0.38 | 9 | 2.18 | 43.496 | 42.817 | 0.967 | 356.096 | 5.27 | 5.42 |
| 79360 | 2012 March 21 01.14416 | 2456007.54767 | 0.12 | 9 | 2.28 | 43.496 | 42.830 | 0.983 | 356.201 | 5.07 | 5.22 |
| 79360 | 2012 March 22 01.05568 | 2456008.54399 | 0.13 | 9 | 2.17 | 43.496 | 42.843 | 0.998 | 356.307 | 5.01 | 5.16 |

Note —Table 2 is published in its entirety in the electronic edition of the Astronomical Journal. A portion is shown here for guidance regarding its form and content.

[a] Mid-time of exposure sequence (calendar date is the same as mid-time in readable format for convenience).

[b] Range of object in Sloan_r' or Bessel R magnitude space during observations on each night.

[c] Duration of observations per night.

[d] $H_{r',Bowell}$ is the absolute magnitude calculated using the formalism of Bowell et al. (1989):

$H = H_{obs}(\alpha) - 5\log(r\Delta) + 2.5\log\left[(1-G)\Phi_1(\alpha) + G\Phi_2(\alpha)\right]$, where $G = 0.15$, $\Phi_1(\alpha) = e^{\left[-3.33\tan(0.5\alpha)^{0.63}\right]}$, and $\Phi_2(\alpha) = e^{\left[-1.87\tan(0.5\alpha)^{1.22}\right]}$.

[e] $H_{r'}(\alpha)$ is the reduced magnitude at unit distance from the Sun and the Earth, uncorrected for phase angle: $H_{r'}(\alpha) = R_{obs}(\alpha) - 5\log(r\Delta)$

TABLE 3. ORBIT INFORMATION AND LIGHTCURVE RESULTS

| Number | Designation | Class[a] | <i>[b] (°) | $H_{r',Bowell}$ | Single/Double Peaked Period (hr) | Peak-Peak Amplitude (mag) | Upper Limit on Amplitude (mag) | # Nights Observed | Notes, Other possible periods (hr) |
|---|---|---|---|---|---|---|---|---|---|
| 79360 | 1997 $CS_{29}$ | CL | 3.84 | 5.09±0.09 | — | — | <0.17 | 10 | Binary, data consistent with a lightcurve period equal to the mutual orbit period, 150.12/300.24 hours |
| 119951 | 2002 $KX_{14}$ | CL | 2.82 | 4.46±0.01 | — | — | <0.05 | 3 | — |
| 120178 | 2003 $OP_{32}$ | SD | 27.02 | 3.82±0.01 | 4.85/9.71 | 0.18±0.01 | — | 6 | Haumea, 6.09/12.18 |
| 120347 | 2004 $SB_{60}$ | SD | 25.57 | 3.97±0.04 | — | — | <0.04 | 4 | Binary, Haumea |
| 145452 | 2005 $RN_{43}$ | SD | 19.45 | 3.47±0.01 | 6.95/13.89 | 0.06±0.01 | — | 4 | 9.73/19.46 |
| 145453 | 2005 $RR_{43}$ | SD | 27.03 | 3.86±0.01 | — | — | <0.06 | 5 | Haumea |
| 174567 | 2003 $MW_{12}$ | SD | 21.24 | 3.18±0.02 | — | — | <0.04 | 4 | Binary |
| 225088 | 2007 $OR_{10}$ | 10:3 | 34.72 | 1.76±0.01 | — | — | <0.09 | 7 | — |
| 278361 | 2007 $JJ_{43}$ | SD | 13.45 | 4.17±0.20 | 6.04/12.09 | 0.13±0.02 | — | 6 | 4.83/9.66 |
| 303712 | 2005 $PR_{21}$ | CL | 2.69 | 5.96±0.02 | — | — | <0.28 | 2 | Binary |
| 303775 | 2005 $QU_{182}$ | SD | 12.82 | 3.36±0.05 | 9.61/19.22 | 0.12±0.02 | — | 6 | — |
| 305543 | 2008 $QY_{40}$ | SD | 24.02 | 5.19±0.06 | — | — | <0.15 | 4 | — |
| 307251 | 2002 $KW_{14}$ | SD | 8.48 | 5.50±0.03 | 6.63/13.25 | 0.25±0.03 | — | 5 | 5.23/10.46 |
| 312645 | 2010 $EP_{65}$ | 2:1 | 19.32 | 5.23±0.15 | 7.48/14.97 | 0.17±0.03 | — | 7 | 5.77/11.54; 8.45/16.90 |
| — | 2005 $EF_{298}$ | CL | 1.60 | 5.73±0.05 | 4.82/9.65 | 0.31±0.04 | — | 3 | Binary, 6.06/12.13 |
| — | 2006 $HJ_{123}$ | 3:2 | 15.62 | 5.92±0.03 | — | — | <0.13 | 4 | — |
| — | 2007 $JF_{43}$ | 3:2 | 14.90 | 5.24±0.05 | 4.76/9.52 | 0.22±0.02 | — | 4 | — |
| — | 2007 $JH_{43}$ | SD | 15.09 | 4.49±0.05 | — | — | <0.08 | 6 | — |
| — | 2009 $YE_{7}$ | SD | 28.00 | 4.26±0.01 | — | — | <0.20 | 4 | Haumea, multiple periods |
| — | 2010 $EK_{139}$ | 7:2 | 29.13 | 3.89±0.04 | 3.53/7.07 | 0.12±0.02 | — | 7 | — |
| — | 2010 $EL_{139}$ | 3:2 | 23.99 | 5.23±0.04 | 3.16/6.32 | 0.15±0.03 | — | 6 | — |
| — | 2010 $ER_{65}$ | SD | 21.58 | 4.69±0.07 | — | — | <0.16 | 4 | — |
| — | 2010 $ET_{65}$ | SD | 30.00 | 4.96±0.11 | 3.94/7.88 | 0.13±0.02 | — | 6 | — |

| Number | Designation | Class[a] | <i>[b] (°) | $H_{r',Bowell}$ | Single/Double Peaked Period (hr) | Peak-Peak Amplitude (mag) | Upper Limit on Amplitude (mag) | # Nights Observed | Notes, Other possible periods (hr) |
|---|---|---|---|---|---|---|---|---|---|
| — | 2010 $FX_{86}$ | SD | 26.60 | 4.34±0.04 | 7.90/15.80 | 0.26±0.04 | — | 8 | — |
| — | 2010 $HE_{79}$ | 3:2 | 15.04 | 5.06±0.04 | 9.75/19.49 | 0.11±0.02 | — | 5 | — |
| — | 2010 $KZ_{39}$ | SD | 25.08 | 4.03±0.01 | — | — | <0.17 | 3 | — |
| — | 2010 $PU_{75}$ | SD | 7.42 | 5.80±0.07 | 6.19/12.39 | 0.27±0.03 | — | 4 | 4.91/9.82 |
| — | 2010 $RF_{43}$ | CN | 34.52 | 3.54±0.04 | — | — | <0.08 | 7 | — |
| — | 2010 $RO_{64}$ | SD | 15.86 | 4.84±0.02 | — | — | <0.16 | 3 | — |
| — | 2010 $TY_{53}$ | CN | 21.56 | 5.34±0.03 | — | — | <0.14 | 4 | — |
| — | 2010 $VK_{201}$ | SD | 27.72 | 4.40±0.07 | 3.79/7.59 | 0.30±0.02 | — | 8 | 3.28/6.55 |
| — | 2010 $VZ_{98}$ | SD | 3.46 | 4.81±0.04 | — | — | <0.18 | 4 | 4.86/9.72, just below significance criterion |

[a] Using the formalism of Elliot et al. (2005): CL = Cold Classical, CN = Centaur, SD = Scattered Disk, M:N = Mean Motion Resonance with Neptune.

[b] The mean heliocentric inclination based on a 10My integration of the orbit of the object; used for identifying an object as Cold Classical.

TABLE 4. INDIVIDUAL MAGNITUDES (ONLINE TABLE)

| Object | Filter | Julian date[a] | $m_{r'}$[b] | $\sigma_{r'}$[b] | r (AU)[c] | Δ (AU)[c] | α (°)[c] | L-Time (s) | $H_{r',Bowell}$[d] | $H_{r'}(\alpha)$[e] |
|---|---|---|---|---|---|---|---|---|---|---|
| 2005EF298 | Sloan_r | 2456005.54549 | 21.927 | 0.027 | 40.655 | 39.715 | 0.469 | 330.299 | 5.791 | 5.887 |
| 2005EF298 | Sloan_r | 2456005.59497 | 21.757 | 0.021 | 40.655 | 39.716 | 0.470 | 330.301 | 5.621 | 5.717 |
| 2005EF298 | Sloan_r | 2456005.59925 | 21.766 | 0.021 | 40.655 | 39.716 | 0.470 | 330.301 | 5.630 | 5.726 |
| 2005EF298 | Sloan_r | 2456005.60353 | 21.804 | 0.022 | 40.655 | 39.716 | 0.470 | 330.301 | 5.668 | 5.764 |
| 2005EF298 | Sloan_r | 2456005.67499 | 21.985 | 0.026 | 40.655 | 39.716 | 0.472 | 330.305 | 5.848 | 5.945 |
| 2005EF298 | Sloan_r | 2456005.67926 | 21.970 | 0.025 | 40.655 | 39.716 | 0.472 | 330.305 | 5.833 | 5.930 |
| 2005EF298 | Sloan_r | 2456005.68354 | 21.965 | 0.024 | 40.655 | 39.716 | 0.472 | 330.305 | 5.828 | 5.925 |
| 2005EF298 | Sloan_r | 2456005.74765 | 22.099 | 0.028 | 40.655 | 39.716 | 0.473 | 330.308 | 5.962 | 6.059 |
| 2005EF298 | Sloan_r | 2456005.75193 | 22.019 | 0.027 | 40.655 | 39.716 | 0.474 | 330.308 | 5.882 | 5.979 |

Note —Table 4 is published in its entirety in the electronic edition of the Astronomical Journal. A portion is shown here for guidance regarding its form and content.

[a] Julian date at the beginning of the exposure, not corrected for light-time.

[b] Apparent sloan r' magnitude and corresponding uncertainty on the individual measurement.

[c] Heliocentric distance, geocentric distance, phase angle, and light time, respectively.

[d] $H_{r',Bowell}$ is the absolute magnitude calculated using the formalism of Bowell et al. (1989):

$H = H_{obs}(\alpha) - 5\log(r\Delta) + 2.5\log\left[(1-G)\Phi_1(\alpha) + G\Phi_2(\alpha)\right]$, where $G = 0.15$, $\Phi_1(\alpha) = e^{\left[-3.33\tan(0.5\alpha)^{0.63}\right]}$, and $\Phi_2(\alpha) = e^{\left[-1.87\tan(0.5\alpha)^{1.22}\right]}$.

[e] $H_{r'}(\alpha)$ is the reduced magnitude at unit distance from the Sun and the Earth, uncorrected for phase angle: $H_{r'}(\alpha) = R_{obs}(\alpha) - 5\log(r\Delta)$

TABLE 5. SPEARMAN'S ρ RANK CORRELATION

| Comparison | All | | | Binary | | | Haumea | | |
|---|---|---|---|---|---|---|---|---|---|
| | $N_{obj}$ | ρ | Sig[a] | $N_{obj}$ | ρ | Sig[a] | $N_{obj}$ | ρ | Sig[a] |
| Amplitude, Aphelion | 128 | -0.143 | 0.108 | 24 | -0.309 | 0.141 | — | — | — |
| Amplitude, Eccentricity | 128 | -0.150 | 0.092 | 24 | -0.293 | 0.165 | — | — | — |
| Amplitude, $H_v$ | 128 | 0.288 | 0.001 | 24 | 0.321 | 0.127 | 9 | 0.092 | 0.814 |
| Amplitude, Inclination | 128 | -0.170 | 0.056 | 24 | -0.319 | 0.128 | — | — | — |
| Amplitude, Perihelion | 128 | 0.036 | 0.690 | 24 | 0.195 | 0.361 | — | — | — |
| Amplitude, Semi-major axis | 128 | -0.094 | 0.291 | 24 | -0.038 | 0.860 | — | — | — |
| Single Period, Aphelion | 69 | 0.249 | 0.039 | 9 | 0.317 | 0.406 | — | — | — |
| Single Period, Eccentricity | 69 | 0.245 | 0.043 | 9 | 0.483 | 0.188 | — | — | — |
| Single Period, $H_v$ | 69 | -0.301 | 0.012 | 9 | -0.267 | 0.488 | 5 | -0.900 | 0.037 |
| Single Period, Inclination | 69 | -0.020 | 0.874 | 9 | 0.283 | 0.460 | — | — | — |
| Single Period, Perihelion | 69 | 0.013 | 0.916 | 9 | -0.233 | 0.546 | — | — | — |
| Single Period, Semi-major axis | 69 | 0.191 | 0.116 | 9 | -0.333 | 0.381 | — | — | — |
| Double Period, Aphelion | 67 | 0.171 | 0.165 | 10 | -0.273 | 0.446 | — | — | — |
| Double Period, Eccentricity | 67 | 0.057 | 0.646 | 10 | -0.103 | 0.776 | — | — | — |
| Double Period, $H_v$ | 67 | -0.143 | 0.249 | 10 | 0.152 | 0.676 | 4 | 0.400 | 0.600 |
| Double Period, Inclination | 67 | -0.075 | 0.546 | 10 | -0.515 | 0.128 | — | — | — |
| Double Period, Perihelion | 67 | 0.115 | 0.354 | 10 | -0.042 | 0.907 | — | — | — |
| Double Period, Semi-major axis | 67 | 0.159 | 0.199 | 10 | -0.552 | 0.098 | — | — | — |

Note – Pluto/Charon and Sila/Nunam are excluded from the rotation period statistics since their rotations are due to tidal locking.

[a] The significance is a value in the interval [0.0, 1.0] where a small value indicates a significant correlation. A 3-sigma result yields a significance of ~0.001.

TABLE 6. POPULATION SAMPLES, K-S TEST

| Sample1 | Sample2 | D | %[a] | N1[b] | N2[b] |
|---|---|---|---|---|---|
| Amplitude non-binary | Amplitude Binary | 0.21 | 69.03 | 104 | 24 |
| Amplitude Classical non-binary | Amplitude Classical binary | 0.21 | 4.04 | 13 | 8 |
| Amplitude Scattered non-binary | Amplitude Scattered binary | 0.31 | 51.78 | 59 | 7 |
| Amplitude Resonant non-binary | Amplitude Resonant binary | 0.22 | 14.21 | 24 | 9 |
| Amplitude Scattered all | Amplitude Haumea all | 0.32 | 67.05 | 71 | 9 |
| Amplitude Scattered all | Amplitude Resonant all | 0.08 | 0.23 | 79 | 32 |
| Amplitude Classical all | Amplitude Scattered all | 0.36 | 95.60 | 17 | 79 |
| Amplitude Classical all | Amplitude Resonant all | 0.39 | 94.64 | 17 | 32 |
| Amplitude Classical all | Amplitude Scattered & Resonant all | 0.36 | 97.05 | 17 | 111 |
| Single period non-binary | Single period binary | 0.23 | 24.46 | 59 | 9 |
| Double period non-binary | Double period binary | 0.27 | 48.37 | 58 | 10 |

[a] The level of confidence that the two groups are not drawn from the same parent population.
[b] The number of objects N1 and N2 used in Sample 1 and Sample 2 respectively.

TABLE 7. PHASE CURVES

| Object | Phase Angle Minimum (°) | Phase Angle Maximum (°) | Phase Angle Range (°) | β (mag/°) | $H_{r'}$ (1,1,0) |
|---|---|---|---|---|---|
| 119951 | 1.17 | 1.46 | 0.30 | 0.13±0.06 | 4.48±0.08 |
| 305543 | 0.60 | 0.95 | 0.36 | 0.42±0.10 | 5.00±0.07 |
| 2010 $FX_{86}$ | 0.60 | 0.99 | 0.39 | 0.23±0.14 | 4.28±0.12 |
| 2010 $EK_{139}$ | 0.63 | 1.04 | 0.41 | 0.19±0.06 | 3.87±0.05 |
| 2010 $EL_{139}$ | 0.54 | 0.97 | 0.44 | 0.05±0.07 | 5.30±0.06 |
| 120178 | 0.43 | 0.87 | 0.44 | 0.13±0.11 | 3.85±0.08 |
| [a]2007 $JF_{43}$ | 0.19 | 1.05 | 0.86 | 0.20±0.05 | 5.22±0.04 |
| [a]2010 $ET_{65}$ | 0.13 | 1.21 | 1.08 | 0.34±0.02 | 4.87±0.01 |
| [a]278361 | 0.22 | 1.32 | 1.09 | 0.56±0.03 | 3.68±0.03 |
| [a]312645 | 0.16 | 1.28 | 1.12 | 0.11±0.03 | 5.34±0.02 |
| [a]2007 $JH_{43}$ | 0.14 | 1.28 | 1.13 | 0.20±0.01 | 4.45±0.01 |

[a] Since α<0.2°, these phase curves could be influenced by opposition surge effects.

TABLE 8. COLORS

| Object | Midtime JD | r' | g' | i' | g'-r' | g'-i' | r'-i' |
|---|---|---|---|---|---|---|---|
| 2006 $HJ_{123}$ | 2455654.68228 | 21.68±0.03 | 22.50±0.08 | 21.09±0.02 | 0.82±0.08 | 1.41±0.08 | 0.59±0.03 |
| 2007 $JF_{43}$ | 2455653.87518 | 21.23±0.03 | 22.19±0.02 | 20.73±0.01 | 0.96±0.03 | 1.46±0.02 | 0.50±0.03 |
| 2007 $JH_{43}$ | 2456063.71009 | 20.49±0.03 | 21.17±0.03 | 20.08±0.01 | 0.68±0.04 | 1.09±0.03 | 0.41±0.03 |
| 2007 $JJ_{43}$ | 2455633.79553 | 20.58±0.03 | 21.35±0.02 | 20.24±0.03 | 0.77±0.03 | 1.12±0.04 | 0.35±0.04 |
| 2008 $QY_{40}$ | 2455836.61329 | 20.96±0.02 | 21.55±0.02 | 20.58±0.03 | 0.59±0.03 | 0.97±0.04 | 0.39±0.04 |
| 2009 $YE_{7}$ | 2455835.80300 | 21.41±0.03 | 21.71±0.02 | 21.13±0.03 | 0.31±0.03 | 0.59±0.03 | 0.28±0.04 |
| 2010 $VK_{201}$ | 2455835.71400 | 21.32±0.12 | 21.81±0.05 | 21.10±0.06 | 0.49±0.13 | 0.71±0.08 | 0.22±0.13 |
| 2010 $EK_{139}$ | 2455629.79418 | 19.88±0.02 | 20.68±0.01 | 19.55±0.01 | 0.80±0.02 | 1.13±0.01 | 0.33±0.02 |
| 2010 $EL_{139}$ | 2455629.73057 | 21.02±0.04 | 21.77±0.11 | 20.60±0.03 | 0.75±0.11 | 1.17±0.11 | 0.42±0.05 |
| 2010 $EP_{65}$ | 2455632.77304 | 20.56±0.03 | 21.42±0.05 | 19.97±0.02 | 0.86±0.06 | 1.45±0.05 | 0.59±0.03 |
| 2010 $ER_{65}$ | 2455631.59008 | 20.52±0.09 | 21.25±0.08 | 20.14±0.08 | 0.73±0.12 | 1.12±0.11 | 0.38±0.12 |
| 2010 $ET_{65}$ | 2455629.68126 | 20.94±0.02 | 21.42±0.03 | 20.28±0.03 | 0.48±0.04 | 1.14±0.04 | 0.66±0.03 |
| 2010 $FX_{86}$ | 2455656.60667 | 20.99±0.02 | 21.82±0.04 | 20.68±0.03 | 0.83±0.05 | 1.14±0.05 | 0.32±0.03 |
| 2010 $HE_{79}$ | 2455653.77667 | 20.70±0.01 | 21.54±0.02 | 20.36±0.01 | 0.84±0.02 | 1.18±0.02 | 0.34±0.01 |
| 2010 $RF_{43}$ | 2455833.55508 | 20.78±0.01 | 21.58±0.03 | 20.38±0.02 | 0.80±0.03 | 1.21±0.04 | 0.41±0.03 |
| 2010 $TY_{53}$ | 2455855.75690 | 20.65±0.02 | 21.24±0.01 | 20.48±0.02 | 0.60±0.02 | 0.77±0.02 | 0.17±0.03 |
| 2010 $VZ_{98}$ | 2455855.80814 | 20.55±0.01 | 21.39±0.03 | 20.02±0.01 | 0.85±0.04 | 1.37±0.04 | 0.53±0.02 |

TABLE 9. LIGHTCURVE COMPILATION (ONLINE TABLE)

| Num | Name | Desig | Class | <i> | s sign | Single Peak (h) | S err | d sign | Double Peak (h) | D err | A sign | Amp-litude | A err | Abs-olute | B | H | semi | ecc | inc | peri-helion | ap-helion | Ref |
|---|---|---|---|---|---|---|---|---|---|---|---|---|---|---|---|---|---|---|---|---|---|---|
| 225088 | X | 2007 OR10 | 10:3E | 34.72 | eq | 0 | 0 | eq | 0 | 0 | eq | 0.09 | 0.02 | 2 | n | n | 67.037 | 0.501 | 30.765 | 33.451 | 100.623 | This work |
| 119979 | X | 2002 WC19 | 2:1E | 7.6 | eq | 0 | 0 | eq | 0 | 0 | lt | 0.03 | 0 | 5.1 | y | n | 48.14 | 0.263 | 9.177 | 35.479 | 60.801 | S07 |
| 312645 | X | 2010 EP65 | 2:1E | 19.32 | eq | 7.48 | 0 | eq | 14.97 | 0 | eq | 0.17 | 0.02 | 5.6 | n | n | 47.426 | 0.303 | 18.922 | 33.056 | 61.796 | This work |
| 26308 | X | 1998 SM165 | 2:1E+ 6:3EI | 13.07 | eq | 3.983 | 0 | eq | 7.1 | 0.1 | eq | 0.56 | 0.03 | 5.8 | y | n | 48.004 | 0.373 | 13.477 | 30.099 | 65.909 | SJ02 |

Column detailed definitions: (1) MPC number; X = not numbered. (2) MPC name; X = not named. (3) MPC preliminary designation. (4) Dynamical classification. (5) Mean inclination over 10 My integration. (6) Sign for single peak period value. (7) Single peak interpretation. (8) Uncertainty in single peak value. (9) Sign for double peak period value. (10) Double peak interpretation. (11) Uncertainty in double peak value. (12) Sign for amplitude value. (13) Amplitude of light curve or variation. (14) Uncertainty in amplitude. (15) Absolute MPC $H_V$ magnitude. (16) Known binary; yes or no. (17) Identified Haumea family object; yes or no. (18) Semi-major axis of heliocentric orbit. (19) Eccentricity of heliocentric orbit. (20) Inclination of heliocentric orbit. (21) Perihelion of heliocentric orbit. (22) Aphelion of heliocentric orbit. (23) Shortened reference or combination of references; last name initial of author (or combination of initials) and year.

Note: A value of "0" has been used if the information is unknown as a way to easily exclude or extract values while selecting samples.

## FIGURES

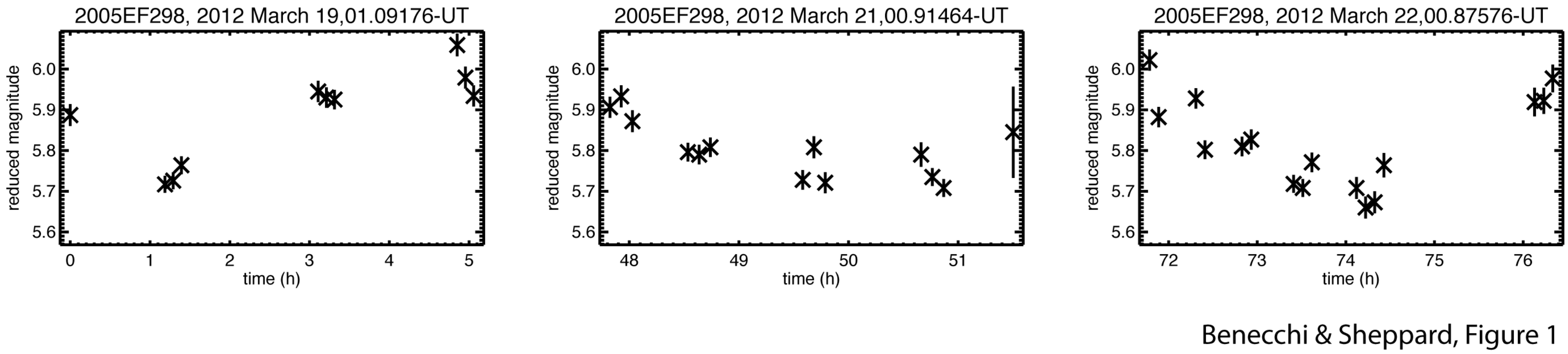

Figure 1. Sample of online plot for each object on each night, in this case for 2005 $EF_{298}$. The y-axis is the calculated $H_{r'}(\alpha)$ magnitude of the object and the x-axis is time in hours referenced to the first observation for this object. The title of each plot is the name of the object and the UT time of the first data point on each individual night of observation.

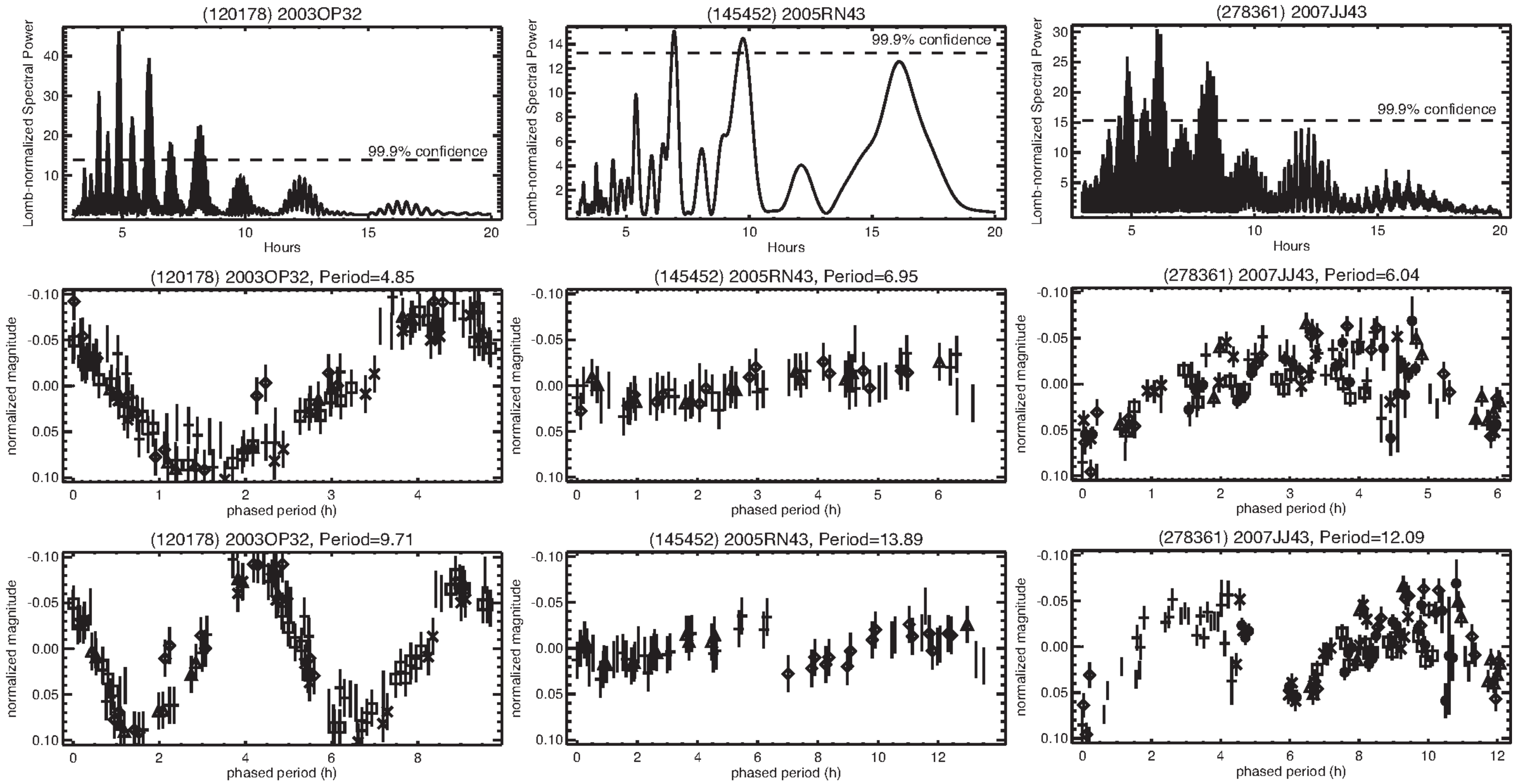

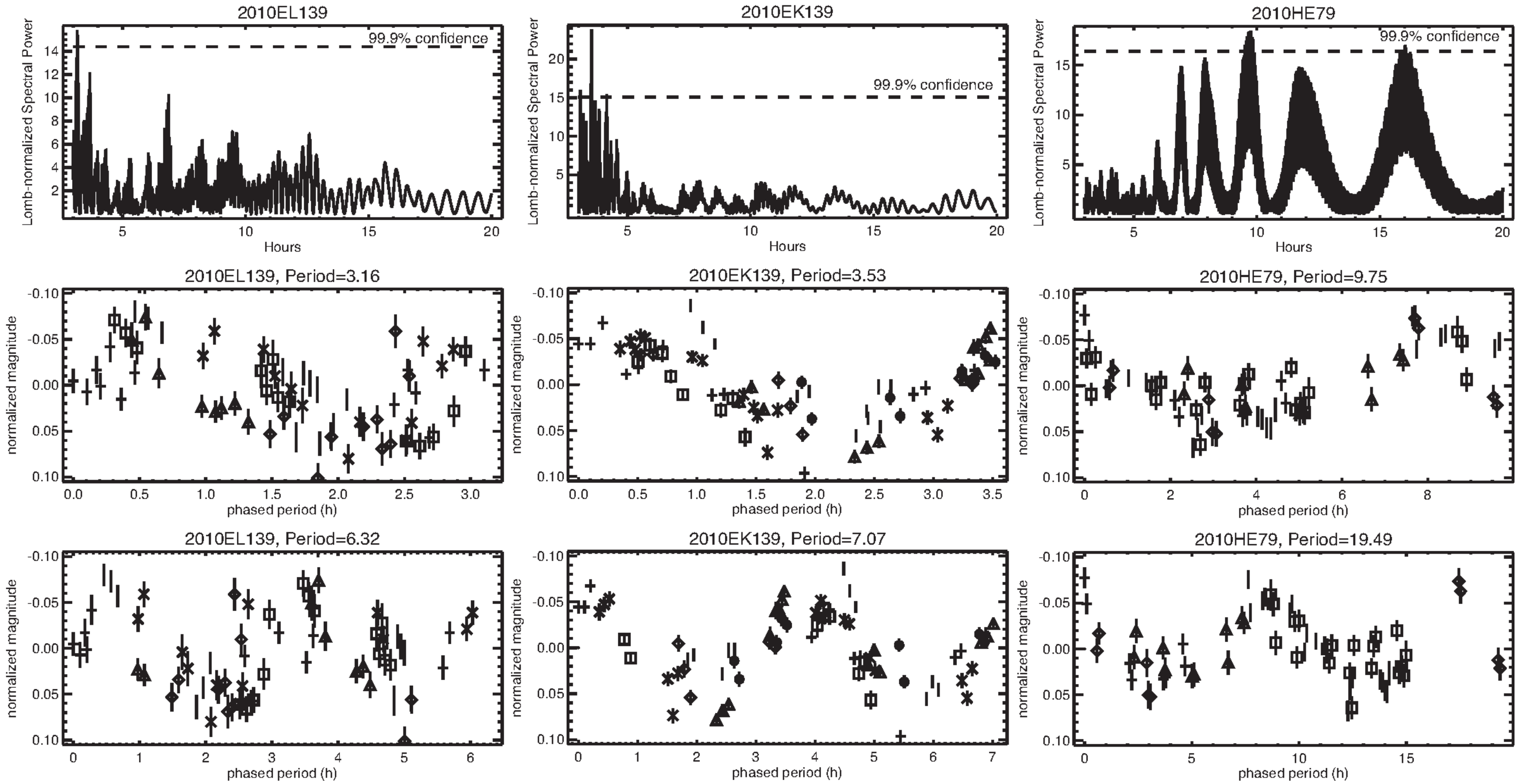

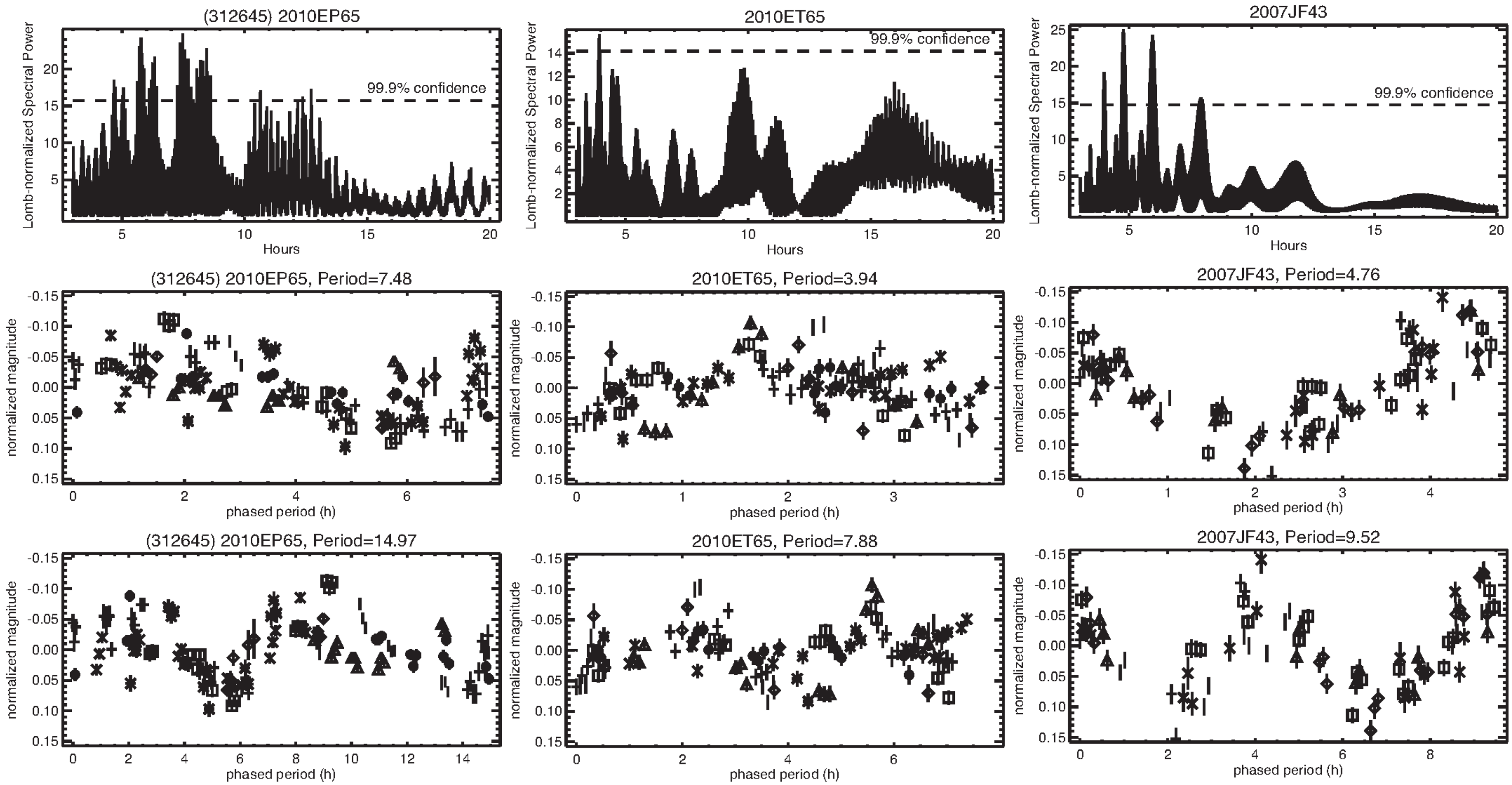

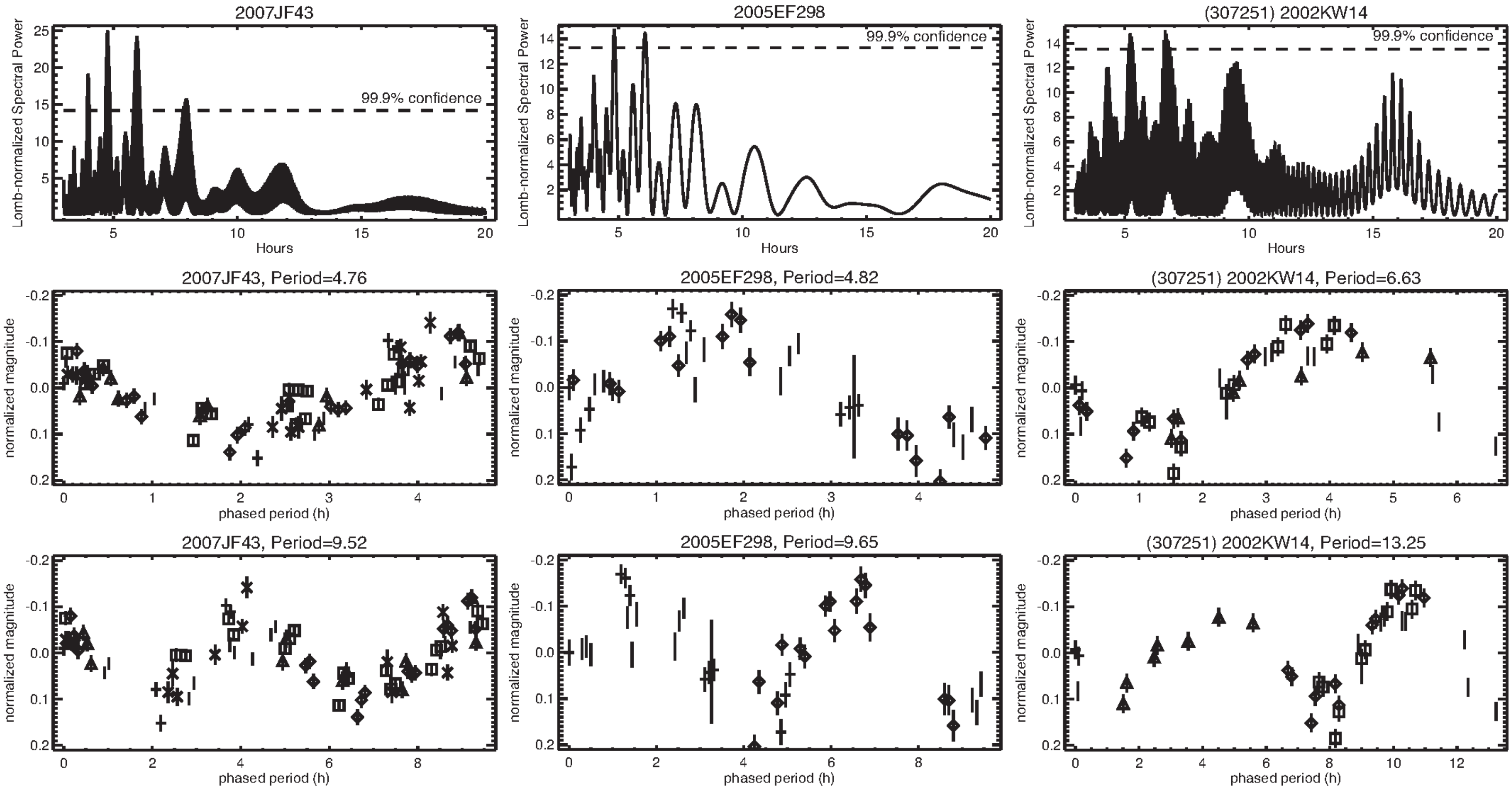

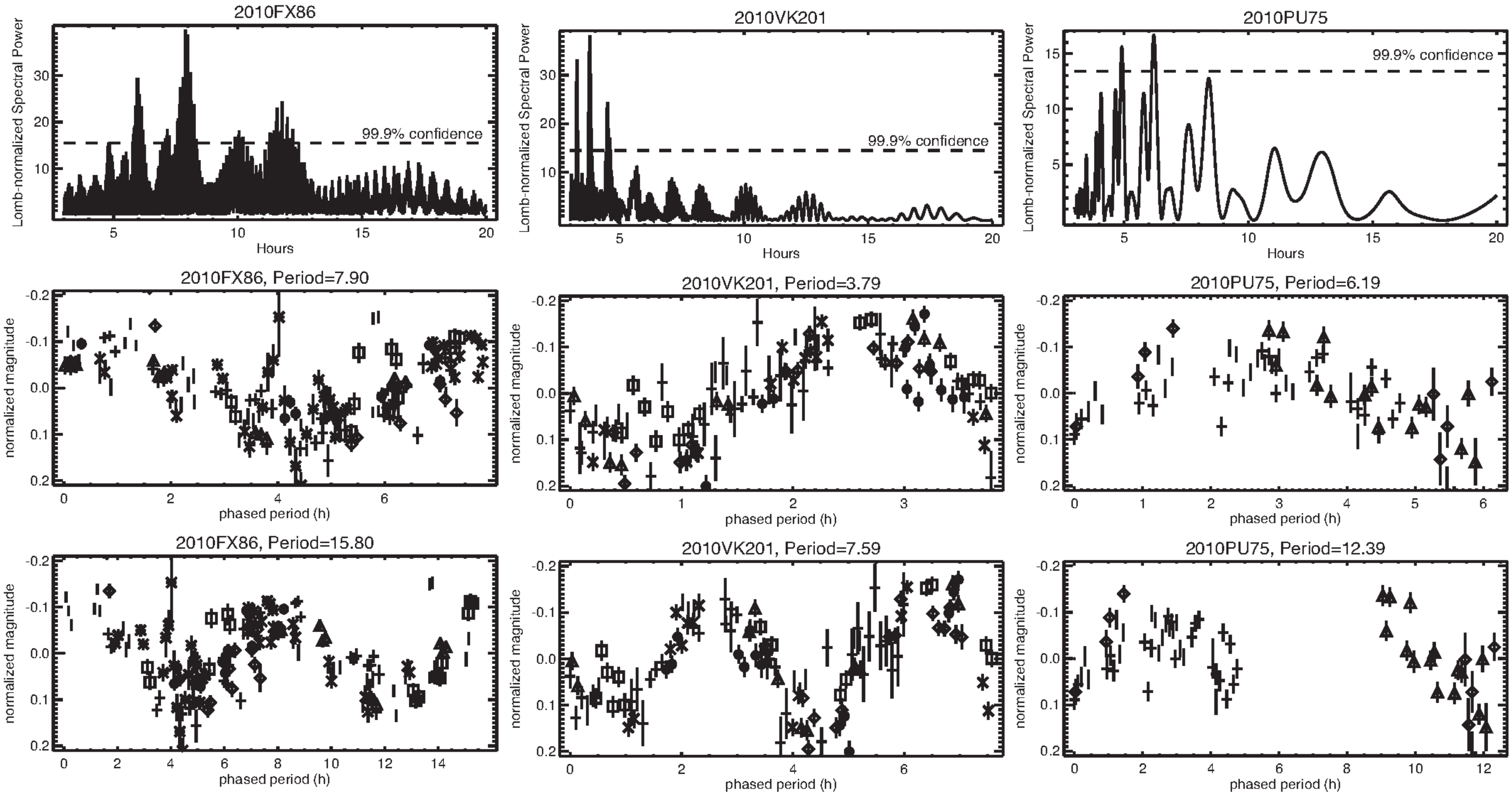

Figure 2. Lightcurve results for objects in this work that have period fits above the 99.9% confidence level. The top panel in each is the Lomb-Scargle periodogram, the middle panel is the data phased to the single-peaked lightcurve period and the bottom panel the data phased to the double-peaked lightcurve period (twice the single-peak interpretation). All periods are given in hours, different symbols show data on different nights.

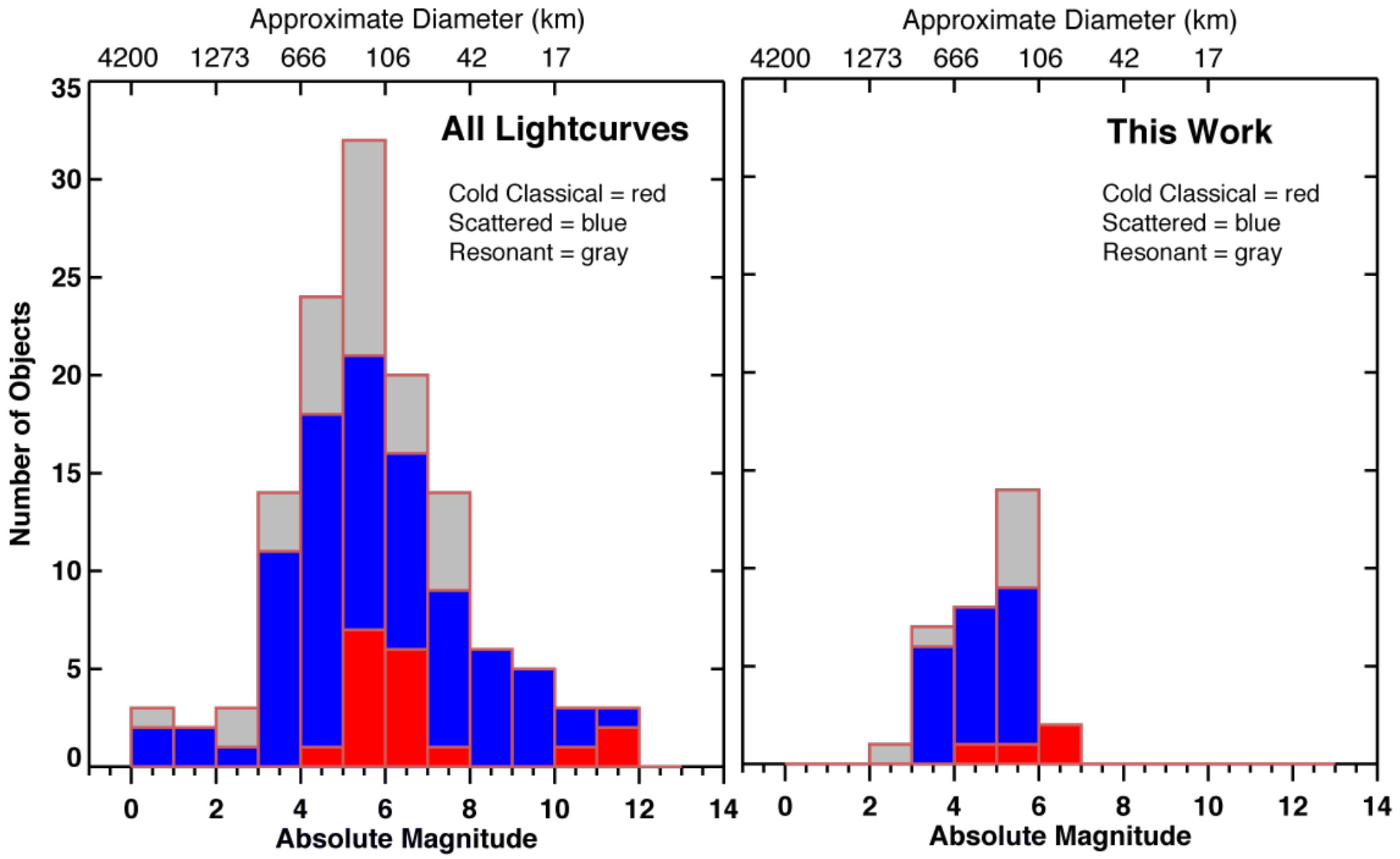

Figure 3. (left) Histogram of all published TNOs observed for lightcurves with respect to the absolute magnitude, $H_V$ (taken from the MPC database for consistency), which is used as a proxy for size; the figure on the right highlights the sample from this work. One can estimate effective diameter from this information following the formulation of Bowell et al. (1989) $d = 10^{((6.259-0.4*H_V-\log\rho)/2)}$ km and assuming an albedo, ρ; we assume ρ=0.1. In addition, objects have been categorized by broad dynamical classification: Cold Classical TNOs (inclinations <5°) are red, Scattered TNOs (inclusive of Scattered objects, Detached objects and Centaurs) are blue, and Resonant TNOs are gray. Original references for compiled datapoints in Figures 3-9: Bus et al. (1989), Tholen & Buie (1990), Buie & Bus (1992), Hoffmann et al. (1992), Buie et al. (1997), Tegler et al. (1997), Davies et al. (1998a), Davies et al. (1998b), Luu & Jewitt (1998), Collander-Brown et al. (1999), Romanishin & Tegler (1999), Consolmagno et al. (2000), Hainaut et al. (2000), Kern et al. (2000), Collander-Brown et al. (2001), Davies et al. (2001), Farnham (2001), Gutierrez et al. (2001), Romanishin et al. (2001), Bauer et al. (2002), Peixinho et al. (2002), Schaefer & Rabinowitz (2002), Sheppard & Jewitt (2002), Bauer et al. (2003), Choi et al. (2003), Farnham & Davies (2003), Ortiz et al. (2003a), Ortiz et al. (2003b), Osip et al. (2003), Rousselot et al. (2003), Sheppard & Jewitt (2003), Chorney & Kavelaars (2004), Mueller et al. (2004), Ortiz et al. (2004), Sheppard & Jewitt (2004), Gaudi et al. (2005), Rousselot et al. (2005a), Rousselot et al. (2005b), Tegler et al. (2005), Belskaya et al. (2006), Kern (2006), Kern & Elliot (2006), Lacerda & Luu (2006), Ortiz et al. (2006), Rabinowitz et al. (2006), Trilling & Bernstein (2006), Lin et al. (2007), Ortiz et al. (2007), Rabinowitz et al. (2007), Sheppard (2007), Dotto et al. (2008), Duffard et al. (2008), Lacerda et al. (2008), Moullet et al. (2008),

Rabinowitz et al. (2008), Roe et al. (2008), Thirouin et al. (2010), Thirouin et al. (2012), and This Work. See online supplemental materials for specific object details.

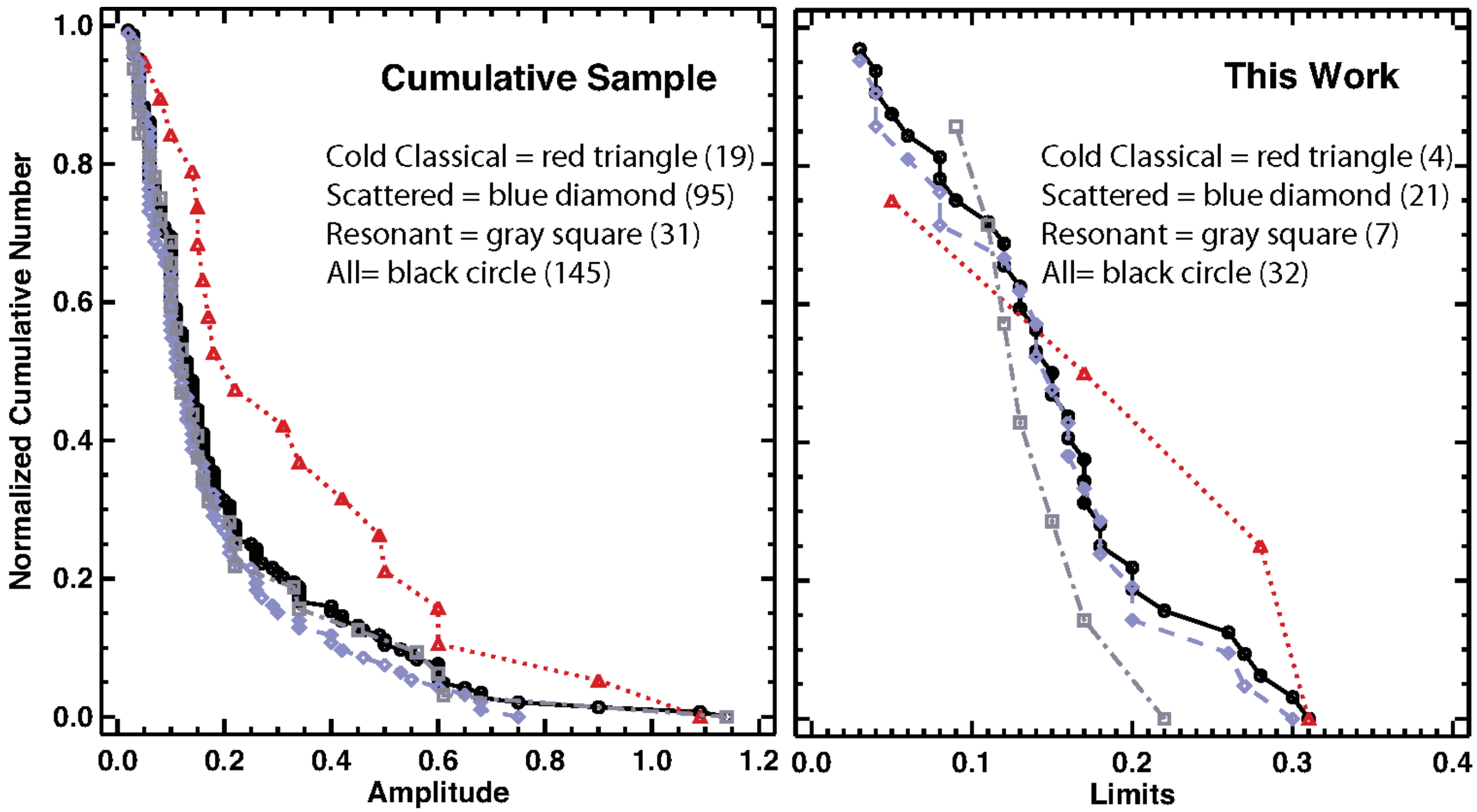

Figure 4. Cumulative number of TNOs amplitudes (or amplitude limits) normalized to the total number of objects in each dynamical sample: (left) all TNOs, and (right) this work only. The black curve is for all TNOs combined, the other samples are as defined such that Cold Classical objects are red triangles, Scattered objects are blue diamonds and Resonant objects are gray squares. The number in parenthesis in the legend is the number of objects in the sample for that panel. The Scattered population has systematically smaller lightcurve amplitudes than the Classical and Resonant populations; this is most evident in the larger sample, although the Resonant objects are dominated by the contact binary 2001 $QG_{298}$. However, Scattered objects at all sizes are measured, so size is not the sole explanation for this effect.

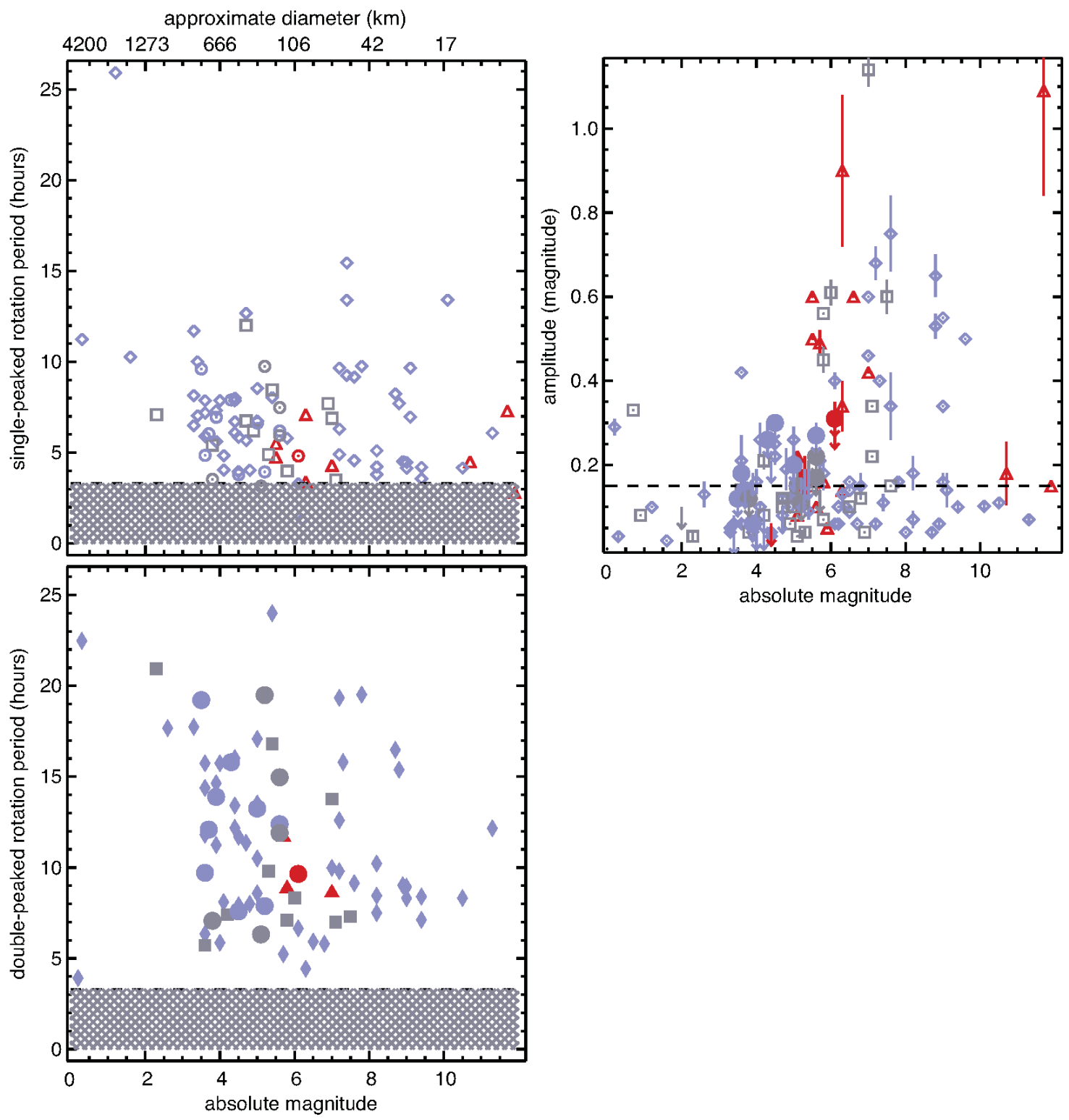

Figure 5. (right) Plot of lightcurve amplitude vs. size (as estimated by $H_V$ magnitude) including objects with only upper limits on their lightcurve amplitudes. The colors and symbols are the same as in Figure 4; solid circles identify the results of this work while open points show results from the literature. A dashed line is drawn at an amplitude of 0.15 magnitudes which is a possible break for interpreting lightcurve amplitudes as being due primarily to surface albedo variations (≤0.15 magnitudes) or due to object elongation [>0.15 magnitudes; Sheppard et al. 2008; Thirouin et al. 2010]. (left) Plot of rotation period (single and double-peaked interpretations) vs. size (as estimated by $H_V$ magnitude) for all objects with published periods. Open symbols indicate single-peaked rotation period interpretation while solid symbols indicate double peaked period interpretation; some objects do not have unique interpretations so both results are plotted. This work's results are plotted with circles. The area below a rotation period of 3.3 hours is grayed out since faster rotation rates would result in the objects being gravitationally unstable assuming a composition of pure ice (Romanishin & Tegler, 1999).

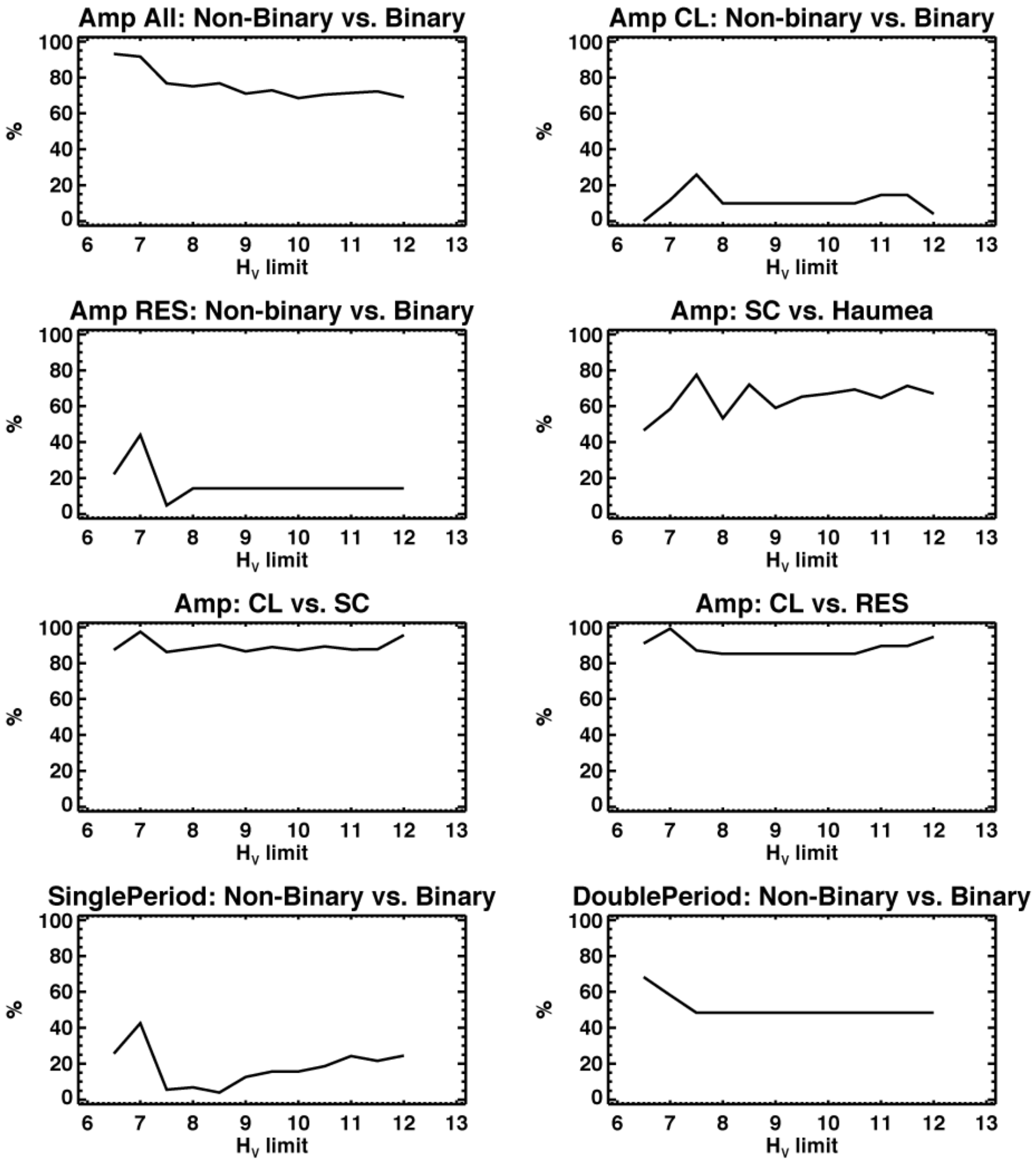

Figure 6. Summary results of KS-test of absolute magnitude limits vs. lightcurve properties in bins of 0.5 magnitudes for each of the samples in Table 5. The amplitude distinction we find is strongest for all three samples (Classical/Resonant; Classical/Scattered and Classical/Scattered+Resonant) with an absolute magnitude limit brighter than 7.0, however, it is strong in all magnitude bins and we believe that the effect is real, not size-dependent.

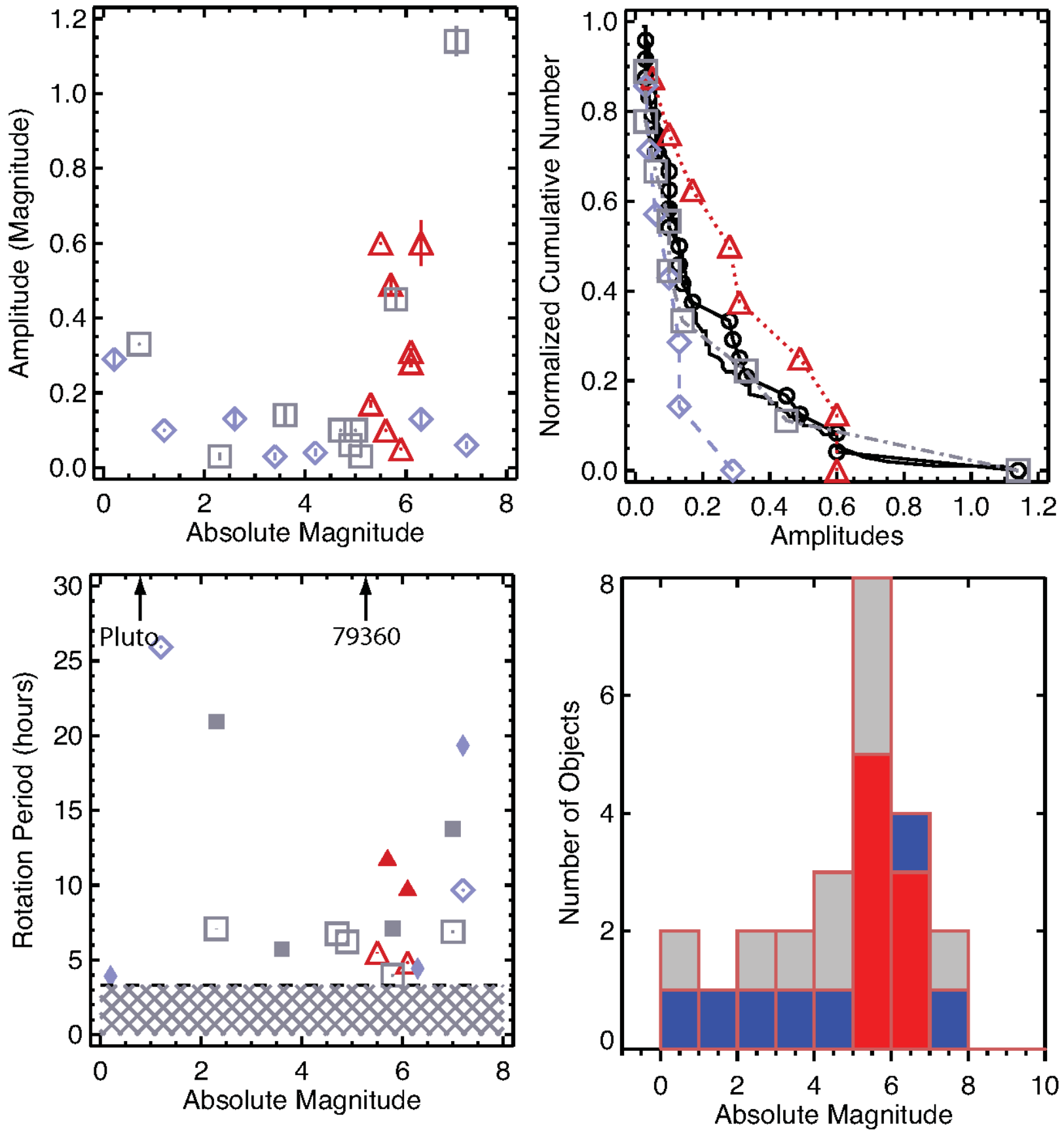

Figure 7. Inventory of lightcurve properties for 24 TNO binaries using the same symbol definitions as in Figures 4. The solid line in the upper right plot is the cumulative curve from Figure 5 for comparison. The binaries show similar lightcurve properties to the larger TNO population, although there is some indication that their lightcurve amplitudes might be, on average, larger. A few binaries, (134340) Pluto/Charon and (79360) Sila/Nunam are tidally locked having lightcurve rotation periods consistent with the mutual orbit period of the system. References: Bus et al. (1989), Tholen & Buie (1990), Romanishin et al. (2001), Sheppard & Jewitt (2002), Ortiz et al. (2003b), Osip et al. (2003), Sheppard & Jewitt (2003), Sheppard & Jewitt (2004), Kern (2006), Kern & Elliot (2006), Lacerda & Luu (2006), Ortiz et al. (2006), Sheppard (2007), Lacerda et al. (2008), Rabinowitz et al. (2008), Thirouin et al. (2012), and this work.

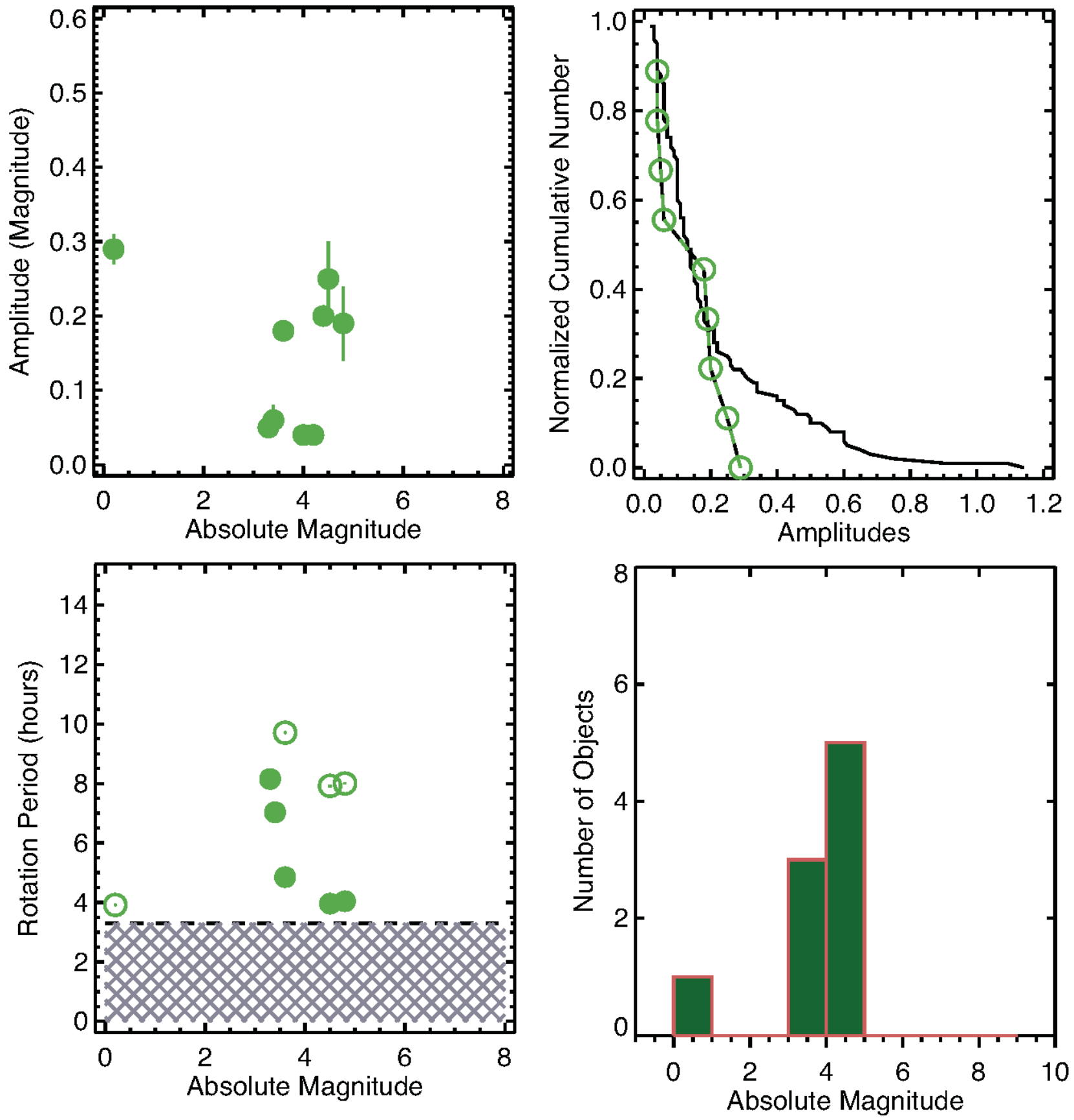

Figure 8. Inventory of lightcurve properties for 9 Haumea family objects. The solid line in the upper right plot is the cumulative curve from Figure 5 for comparison. In the lower left plot solid circles are single-peaked period interpretations and open circles are double-peak period interpretations. All objects have rotation periods (where measured) ≤10 hr and distinguishable light curve amplitudes. At this point, only the brightest Haumea family objects have been observed. Since most of the amplitudes are small, it is likely that they can be interpreted as spherical objects with surface variations due to albedo features. References: Sheppard & Jewitt (2002), Sheppard & Jewitt (2003), Lacerda et al. (2008), Thirouin et al. (2012), and this work.

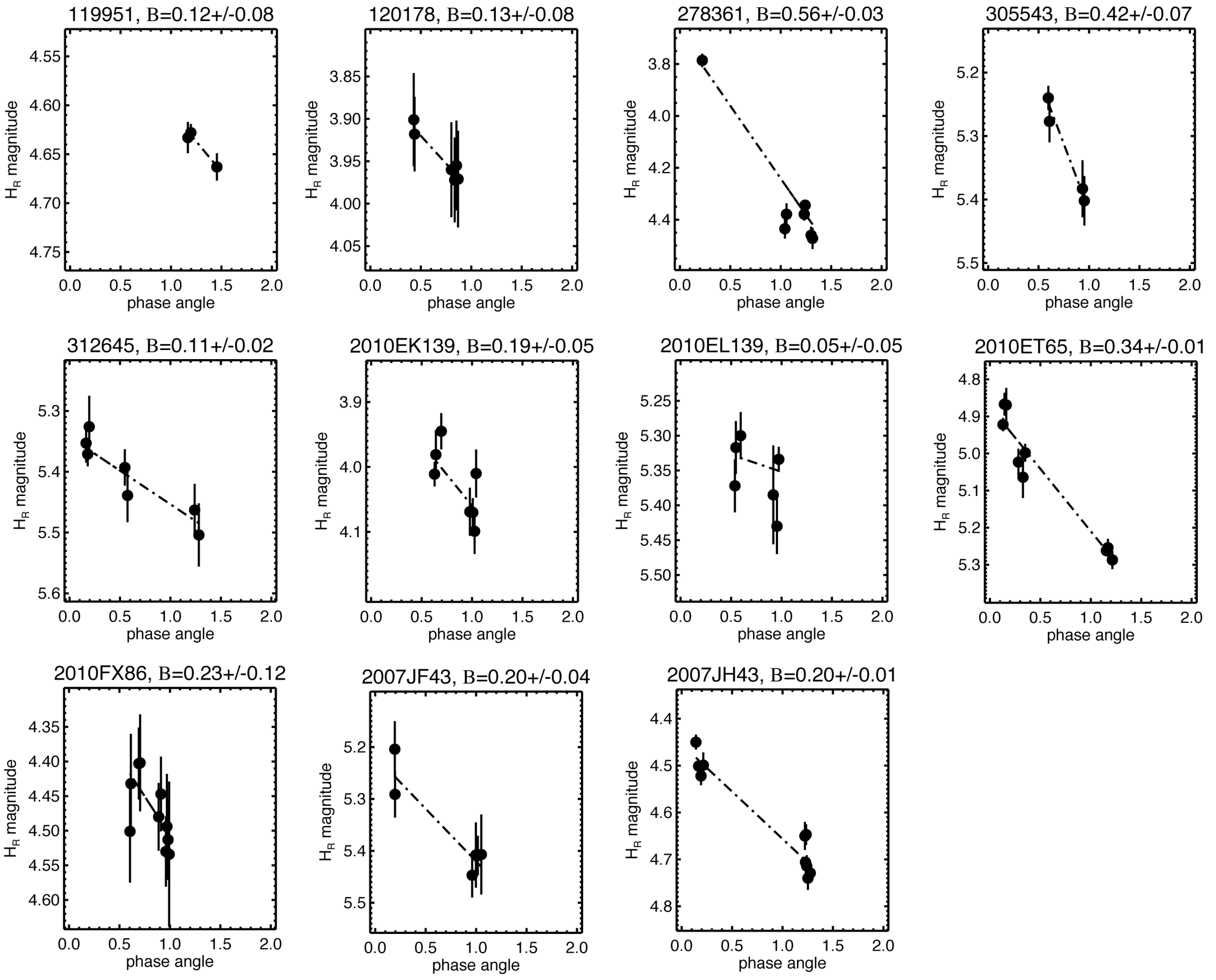

Benecchi & Sheppard, Figure 9

Figure 9. Phase curves for objects in our sample with phase observations ≥0.3°. Most of our objects have phase values similar to those found of other TNO studies (Schaefer et al. 2009). However, (278361) 2007 $JJ_{43}$ has a steep slope and relatively small scatter. It may be that this object displays strong opposition effects since it was observed at a phase angle ≤0.2°.

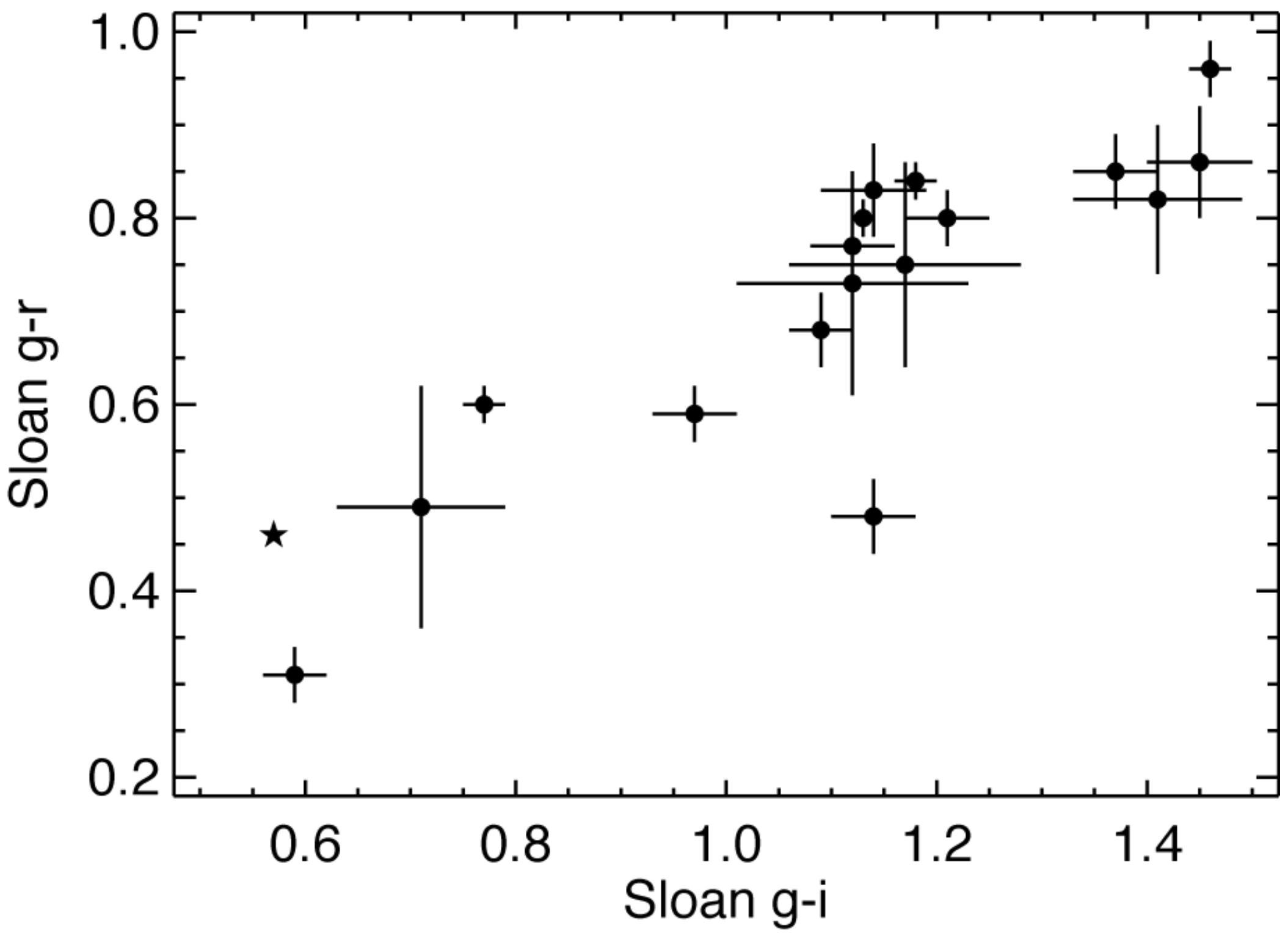

Figure 10. Sloan colors of some of the objects in our survey. All of our objects are redder than the Sun (indicated by a star) with the exception of 2009 $YE_7$ (at ~0.6,0.3), which is one of the Haumea family members. The colors occupy a large range similar to that found for other TNO color surveys.